\documentclass[12pt,preprint]{aastex}

\begin{document}

\title{\large \sc Constraining Disk Parameters of Be Stars using
Narrowband H$\alpha$ Interferometry with the NPOI}

\author{
Christopher Tycner,\altaffilmark{1,2,3} 
G.~C.~Gilbreath,\altaffilmark{4}
R.~T.~Zavala,\altaffilmark{2}
J.~T.~Armstrong,\altaffilmark{4}
J.~A.~Benson,\altaffilmark{2}
Arsen~R.~Hajian,\altaffilmark{5}
D.~J.~Hutter,\altaffilmark{2}
C.~E.~Jones,\altaffilmark{6}
T.~A.~Pauls,\altaffilmark{4}
N.~M.~White,\altaffilmark{7}\\
{\tt \footnotesize tycner@nofs.navy.mil}\\
{\it Accepted for publication in the Astronomical Journal}
}

\altaffiltext{1}{Michelson Postdoctoral Fellow.}
\altaffiltext{2}{US Naval Observatory, Flagstaff Station, 10391
W.~Naval Observatory Rd., Flagstaff, AZ 86001-8521}
\altaffiltext{3}{NVI, Inc., 7257 Hanover Parkway, Suite D, Greenbelt,
MD 20770} 
\altaffiltext{4}{Remote Sensing Division, Code 7210, Naval Research
Laboratory, 4555 Overlook Avenue, SW, Washington, DC 20375}
\altaffiltext{5}{US Naval Observatory, 3450 Massachusetts Avenue, NW,
Washington, DC 20392-5420} 
\altaffiltext{6}{Department of Physics and Astronomy, The University
of Western Ontario, London, Ontario, N6A~3K7, Canada}
\altaffiltext{7}{Lowell Observatory, 1400 West Mars Hill Road,
Flagstaff, AZ, 86001}

\slugcomment{\small \today}

\begin{abstract}
Interferometric observations of two well-known Be stars, $\gamma$~Cas
and $\phi$~Per, were collected and analyzed to determine the spatial
characteristics of their circumstellar regions.  The observations were
obtained using the Navy Prototype Optical Interferometer equipped with
custom-made narrowband filters.  The filters isolate the H$\alpha$
emission line from the nearby continuum radiation, which results in an
increased contrast between the interferometric signature due to the
H$\alpha$-emitting circumstellar region and the central star.  Because
the narrowband filters do not significantly attenuate the continuum
radiation at wavelengths 50~nm or more away from the line, the
interferometric signal in the H$\alpha$ channel is calibrated with
respect to the continuum channels.  The observations used in this
study represent the highest spatial resolution measurements of the
H$\alpha$-emitting regions of Be stars obtained to date.  These
observations allow us to demonstrate for the first time that the
intensity distribution in the circumstellar region of a Be star cannot
be represented by uniform disk or ring-like structures, whereas a
Gaussian intensity distribution appears to be fully consistent with
our observations.
\end{abstract}


\section{Introduction}

The application of long-baseline optical interferometry to the study
of classical Be stars, although still a developing observational
field, has already resulted in significant contributions to our
understanding of these objects.  This is in part related to the fact
that the circumstellar regions associated with the closest Be stars
can be spatially resolved using optical interferometers with only
modest baseline lengths (10--40~m), especially if the observations are
sensitive to the H$\alpha$ line emission.  The interferometric
observations at such baselines typically yield only angular sizes of
the emitting regions, and in some cases also the apparent ellipticity.
However, when these types of results are combined with polarimetry or
spectroscopy, then a number of different intrinsic properties of these
regions can be investigated or constrained.

For example, \citet{Quirrenbach97} combined optical interferometric
and spectropolarimetric observations of seven Be stars and showed that
the disk orientation inferred from the two completely independent data
sets agree.  Furthermore, because the smallest upper limit on the disk
opening angle derived by \citet{Quirrenbach97} from the apparent
ellipticity of the circumstellar region was $\sim$20$^{\circ}$, this
was the first study that spatially resolved the circumstellar disks
and supported the thin-disk paradigm for Be stars.  \citet{Wood97}
constructed disk models for the Be star $\zeta$~Tau with large (tens
of degrees) and small (few degrees) disk opening angles, where both
models were consistent with the spectropolarimetry.  Using the
interferometric results of \citet{Quirrenbach97} the large opening
angle solution was rejected by \citet{Wood97}.

Another example of the synergy between spectroscopy and long-baseline
interferometry was demonstrated by \citet{Vakili98} when the
observations of the Be star $\zeta$~Tau at two epochs were used to
detect the presence of a one-armed oscillation in the circumstellar
disk.  In a similar study, \citet{Berio99} have shown that the
variable asymmetric brightness distribution in the disk of
$\gamma$~Cas deduced from the optical interferometric data correlate
with the spectral variations seen in the H$\beta$ emission line, and
these in turn can also be explained with a precessing one-armed
oscillation in the equatorial disk.  Although the combination of
spectroscopic and interferometric observations is necessary in studies
related to temporal variability, these types of data sets can also be
combined to put direct constraints on the physical conditions within
the circumstellar regions of Be stars.  For example, \citet{Tycner05}
combined H$\alpha$ emission profiles from spectroscopy with
interferometric observations in H$\alpha$ for a number of different Be
stars and showed that there exists a direct relationship between the
physical extent of the emitting region and the net H$\alpha$
luminosity.

The total number of Be stars investigated with optical interferometry
to date still remains quite small.  This can be attributed to the
relatively small number of long-baseline interferometric instruments
that are sensitive to one of the strongest spectral features produced
within the circumstellar disk, i.e., the H$\alpha$ emission line.
Even the instruments that are sensitive to the H$\alpha$ emission
line, or any other emission line for that matter, and have baselines
long enough to spatially resolve the circumstellar regions, still need
to separate the signatures from the resolved disk and the unresolved
(or nearly unresolved) central star that is detected at the continuum
wavelengths.  For this reason, it is imperative that the instrument be
configured such that the contribution from the stellar photosphere to
the interferometric observations at the H$\alpha$ line is minimized.
This can be accomplished by either sufficiently high spectral
dispersion or with the use of narrowband filters.

The Navy Prototype Optical Interferometer~(NPOI) is by design a
multi-spectral instrument.  However, the continuum bandpass recorded
by the instrument in the H$\alpha$ region, as defined by the spectral
channel characteristics, can be too wide for sources with weak
H$\alpha$ emission.  Therefore, as part of a proof-of-concept, we
equipped the NPOI with a set of custom narrowband filters\footnote{The
custom H$\alpha$ filters were manufactured by David E. Upton of Omega
Optical, Brattleboro, VT.} centered on the H$\alpha$ line to further
decrease the stellar contribution at the continuum wavelengths in the
spectral channel containing the H$\alpha$ emission
line~\citep{Pauls01}.  These initial observations demonstrated that,
with only small modifications to our current instrumental
configuration, interferometric observations of Be stars can be
obtained with the narrowband filters.

In this paper, we demonstrate the quantitative results based on the
narrowband interferometric observations of two stars known for their
relatively strong H$\alpha$ emission, $\gamma$~Cassiopeiae~(=HR~264)
and $\phi$~Persei~(=HR~496).  Because the circumstellar regions of
both stars were resolved by long-baseline interferometry in the past,
they are suitable targets for direct comparison of the results
produced by the new observational setup with those already presented
in the literature~\citep{Quirrenbach97,Tycner03}.  Also, the
interferometric observations presented in this study represent some of
the highest spatial resolution measurements of the H$\alpha$-emitting
regions obtained to date.  This allows us to test the various models
for intensity distribution at high enough spatial frequencies so that
the degeneracy between the various models is eliminated.

\section{Observational Setup}
\label{sec:setup}

\begin{figure}
\plotone{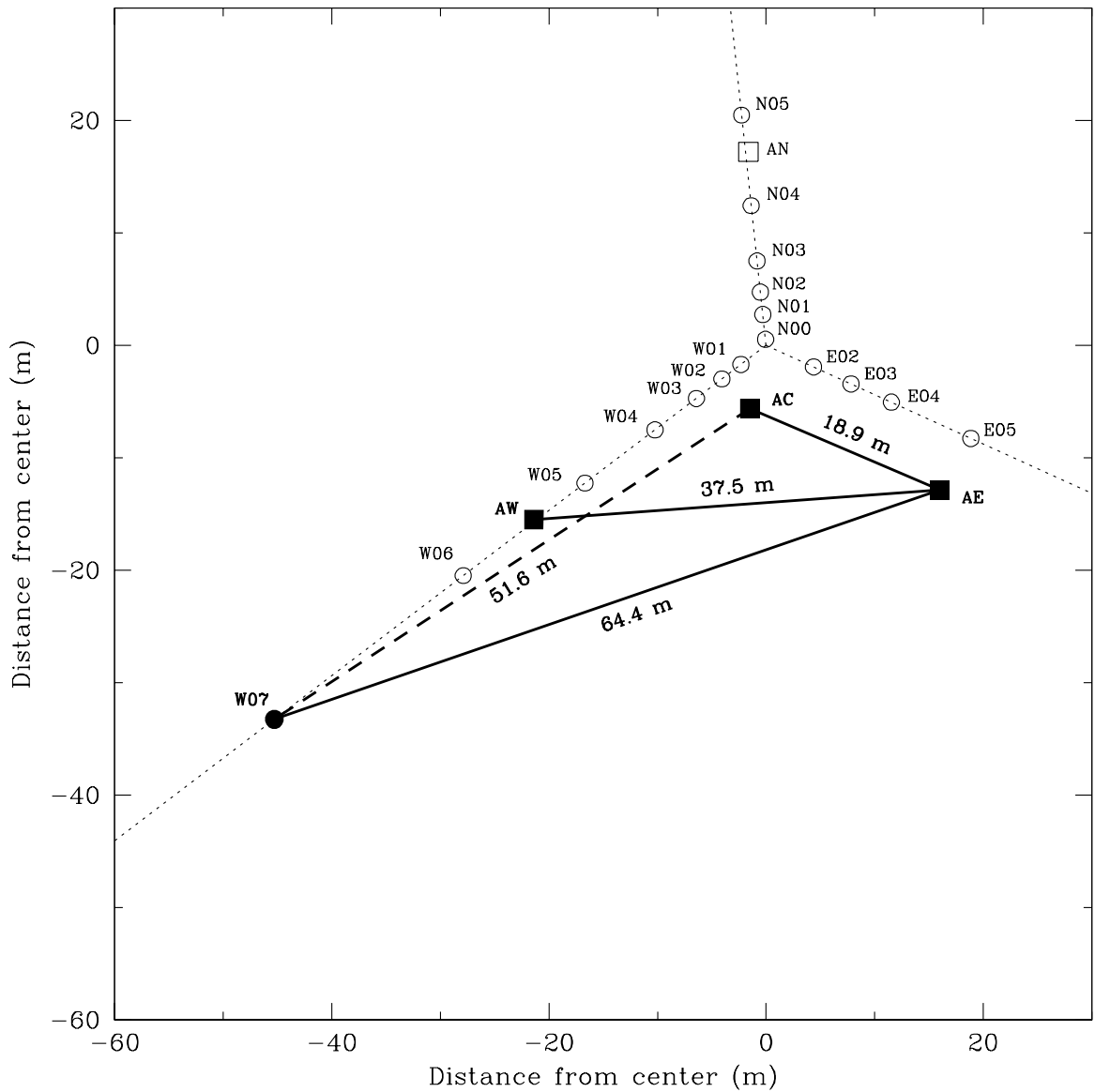}
\caption{Schematic of the inner potion of the NPOI array.  The
imaging~({\it circles}) and astrometric~({\it squares}) stations used
in the H$\alpha$ observations presented in this paper are shown with
filled symbols.  Baselines that could be recorded on two out of three
outputs from the beam combiner are also shown with their respective
physical lengths indicated.  The baseline that was not accessible to
observations of $\gamma$~Cas due to sky coverage limitation is shown
with the dashed line.}
\label{fig:baselines}
\end{figure}

The NPOI consists of four stationary astrometric stations and a
Y-shaped imaging station configuration where movable siderostat
elements can be repositioned~\citep[see the description of the
instrument by][]{Armstrong98}.  Currently, two movable stations are
operational.  The goal is to have six movable stations that will allow
simultaneous observations with up to six reconfigurable elements.
Although the successful combination of light from six stations has
been already demonstrated with the NPOI~\citep{Hummel03}, at the time
of the observations presented in this paper only four stations were
available for observations.  These four stations gave access to
baselines with lengths in the range of 18.9 to 64.4~m, as shown
schematically in Figure~\ref{fig:baselines}.

\begin{figure}
\plotone{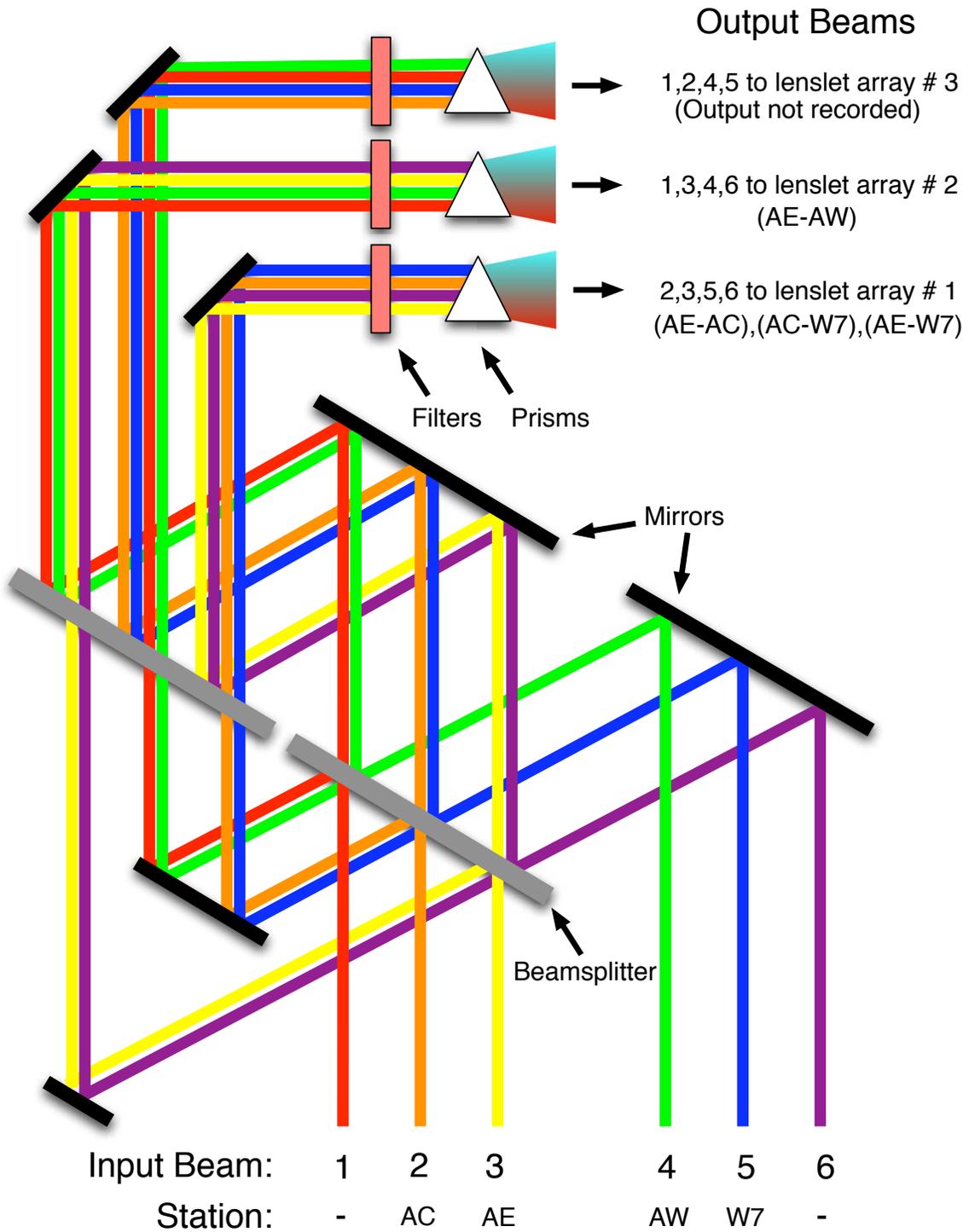}
\caption{Schematic of the beam combiner illustrating the propagation
of light from up to six input beams.  The three output beams that are
intercepted by the pick-off optics sample all available baseline
configurations~(up to 15 element pairs with 6 input beams).  }
\label{fig:beamcombiner}
\end{figure}

The instrument was designed to accept stellar light from up to 6
siderostats simultaneously, where each element sends a 12~cm light
beam through vacuum pipes to the beam combiner lab.  At the beam
combiner, the light beams are split and recombined in such a way that
the interference fringes between all siderostat pairs~(for a total of
15 unique pairs) are constructed.  Figure~\ref{fig:beamcombiner}
illustrates the propagation of light for six input beams.  The three
output beams that are intercepted are dispersed by a set of three
prisms onto individual lenslet arrays, which in turn are fiber-coupled
to a cluster of photon-counting avalanche photodiodes.  Each fiber
corresponds to a single channel with spectral characteristics set by
the position of the lenslet array and the physical width of a single
lenslet along the dispersion direction.

The spectra of Be stars show many spectral lines in emission.  For
example, in the 510--880~nm region covered by our interferometric
observations emission lines of elements such as H~I, He~I, O~I, Si~II,
and Fe~II can be detected.  However, except for hydrogen lines all of
the emission lines are very weak having equivalent widths~(EW) of
$\lesssim$~0.1~nm, and therefore will be lost in the continuum signal
present in each spectral channel~(the widths of the spectral channels
range from 10 to 26~nm).  Even in the case of the hydrogen lines
present in our spectral region of interest only the H$\alpha$ line has
large enough EW to contribute a detectable interferometric signal in
the H$\alpha$ containing channel.  Therefore, for the purpose of the
H$\alpha$ observations, the lenslet array is aligned such that the
emission line is centered on a single channel that has a spectral
width of 15~nm.  Because the typical equivalent widths of H$\alpha$
emission lines in Be stars are usually not more than 4~nm~\citep[see
for example Table 6 in][]{Tycner05}, the contribution from the
circumstellar region to the net signal is typically less than 25~\%.
For sources with very weak emission, the fractional contribution to
the net signal can be much less than the magnitude of the random and
systematic uncertainties, which are typically at the few percent
level.

\begin{figure}
\plotone{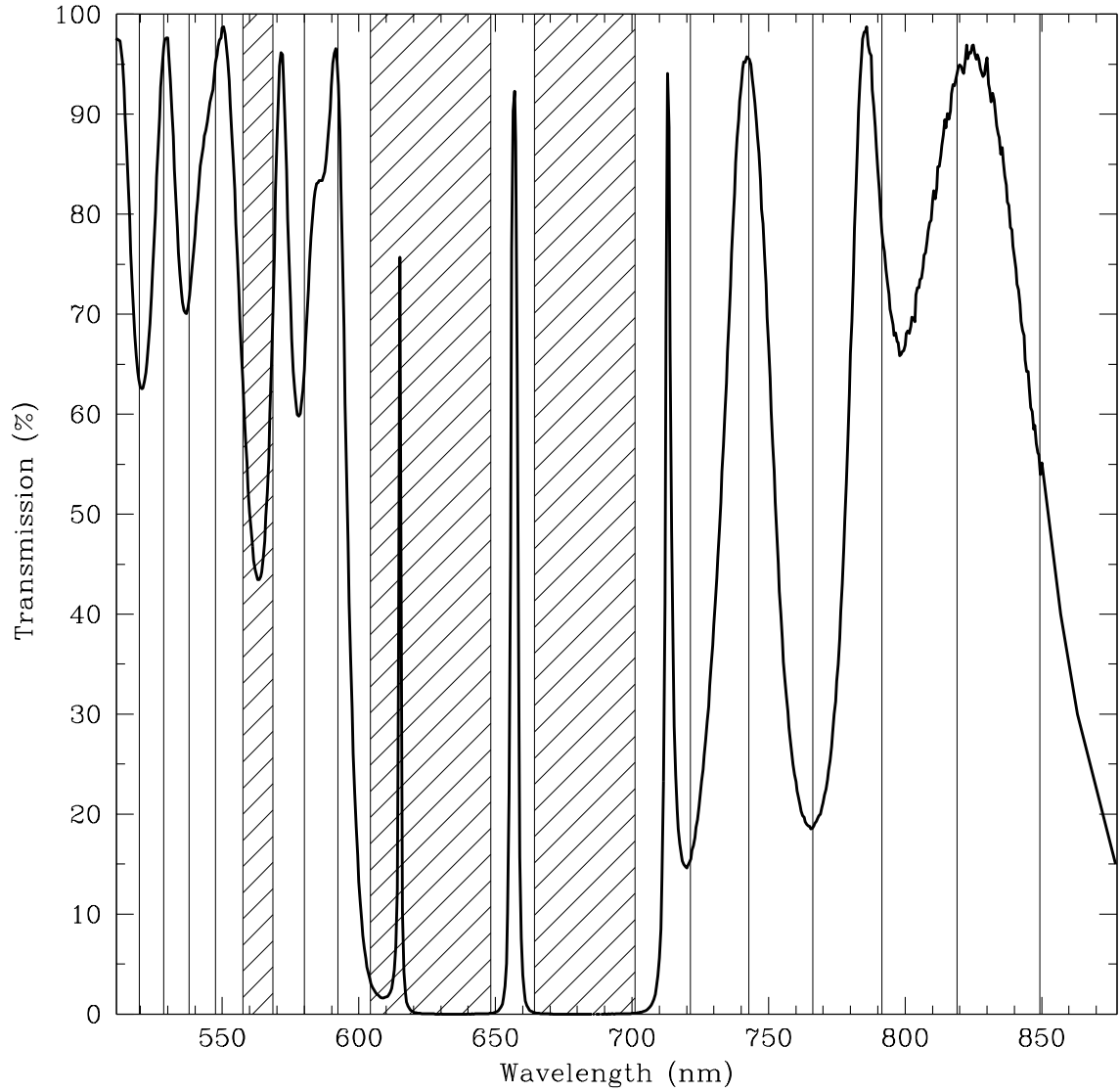}
\caption{Transmission curve for the H$\alpha$ filter in the spectral
region covered by the 16 channels that were used to record the
interferometric signals.  The edges of the spectral channels are
marked with vertical lines, and the spectral regions for which the
signals were not recorded are shown with hatched regions. }
\label{fig:filter}
\end{figure}

To increase the contrast between the interferometric signatures from
the H$\alpha$-emitting circumstellar region and the central star, we
have inserted a narrowband filter at each of the three outputs from
the beam combiner~(see Fig.~\ref{fig:beamcombiner}).  These filters
have a high transmission~($\sim$90~\%) in a 2.8~nm region centered on
the H$\alpha$ line, which suppresses the nearby continuum emission,
but does not significantly attenuate radiation outside a $\sim$100~nm
region centered at the line.  Because we need to detect the
interferometric signal outside of the spectral channel containing the
H$\alpha$ emission, for both calibration and fringe tracking purposes,
the filter has high enough transmission in the 500--600 and
710--850~nm regions that the continuum channels are still usable.  The
transmission curve of the custom-made filter is shown in
Figure~\ref{fig:filter} along with the superposed locations of the
spectral channels that were used to record the fringe signals.

\begin{figure}
\plotone{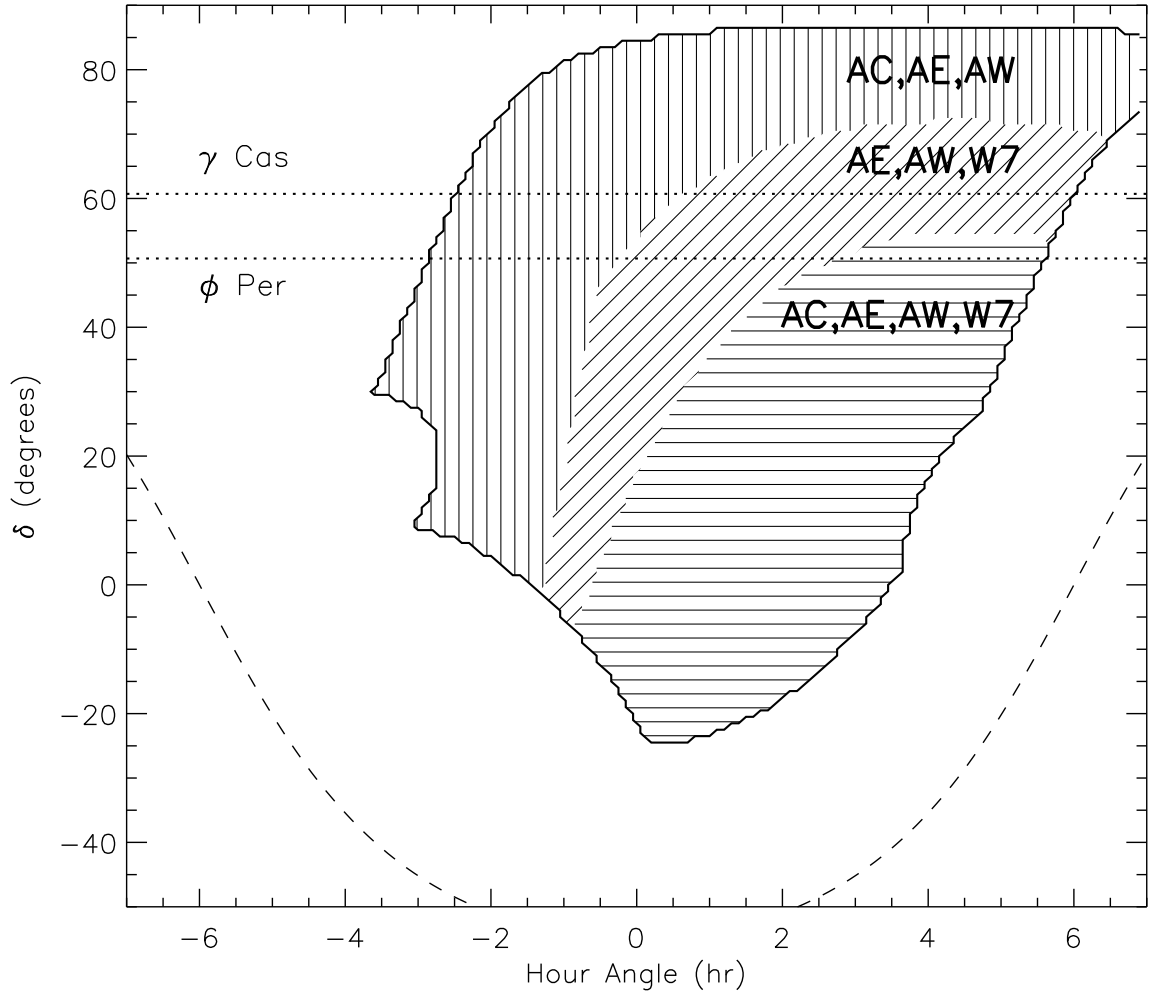}
\caption{Accessible sky coverage for interferometric observations
using different combinations of the array elements.  At a fixed
declination, the largest accessible HA range is determined by the
limits imposed on the siderostats motion~(to the east) and on a
typical zenith angle limit of 60$^{\circ}$~(to the west).  The
accessible HA range is further constrained by limitations imposed by
optical path compensation components.  The smallest area~({\it
horizontal hatched region}) corresponds to observations at all four
stations~(AC, AE, AW, and W7), which has a declination limit of
55$^{\circ}$.  The declination limit as well as the sky coverage can
be increased by excluding the AC~({\it diagonal hatched region}) or
the W7 station~({\it vertical hatched region}).  The positions of the
horizon~({\it dashed line}), $\phi$~Per, and $\gamma$~Cas ({\it dotted
lines}) are also indicated.}
\label{fig:sky}
\end{figure}

In the current instrumental setup the signals from up to 32 channels
can be recorded simultaneously by the NPOI electronics.  Because these
channels are divided equally between two output beams from the beam
combiner, data from one of the three output beams is currently not
recorded~(recall Fig.~\ref{fig:beamcombiner}).  This results in
interferometric observations that do not sample all the baseline
configurations that are possible with a given set of elements, as
shown in Figure~\ref{fig:baselines}.  Furthermore, because the
long-delay lines are not implemented yet, the maximum path difference
that can be introduced between a pair of siderostats~(to compensate
for the different distances between each station and the star) is
currently 35~m and this results in restricted ranges of declination
and hour angle over which a star can be observed at a given baseline.
For example, Figure~\ref{fig:sky} shows the accessible sky coverage
for three different configurations; in one case all four elements (AC,
AW, AE, and W7) are used and in the other two configurations either
the AC or the W7 station is excluded.  Therefore, for $\gamma$~Cas the
observations utilizing the AC--W7 baseline are not currently possible,
whereas for $\phi$~Per, which is 10$^{\circ}$ lower in the sky, such
observations are possible.  Of course, with the expected
implementation of the long delay-lines, optical path difference
compensation of up to $\sim$400~m will become possible and the
accessible sky coverage will be significantly less restrictive.

\section{Observations}

\subsection{Interferometry}
\label{sec:interferometry}

The typical NPOI observing and data reduction procedures are described
in \citet{Hummel03}.  Here we provide a brief review of the process.
During a 30~s observation fringe data are recorded every 2~ms.  The
squared visibility~($V^2$) estimators from the 2~ms data are averaged
on 1~s intervals and the 1~s data points are processed using a suite
of custom reduction software.  Outlier points are flagged on the basis
of the residuals of the delay, seeing indicators, photon rates, and
squared visibilities.  Finally, depending on how many 1~s data points
have been flagged as bad, up to 30 points are averaged to produce a
single squared visibility measure for each 30~s interval, also known
as a ``scan''.  Typically, following a procedure similar to that used
in obtaining the $V^2$ data, the closure phase and triple amplitude
can also be obtained for each scan.  However, because these quantities
were not obtained for the data presented in this paper, we do not
include them in our discussion.

The $V^2$ value for a single scan is obtained in the same way for both
the target and the calibrator star, as well as for the incoherent
scans~(also known as the ``off-the-fringe scans'').  The typical
observational sequence consists of a pair of coherent and incoherent
scans on a target, followed by the same pair of scans on a calibrator.
The procedure is repeated for as long as the scans of the target star
are acquired.  The incoherent scans are necessary to estimate the
additive bias terms affecting the squared visibility
measures~\citep[see][for more details]{Hummel03}.  The bias term is a
squared visibility measure that one obtains in the case of a
completely incoherent signal, which for an ideal system should be
exactly zero.  Because the calibration procedure removes
multiplicative effects, the subtraction of the additive bias term must
be performed before calibration corrections are applied.  We remove
the bias terms from the squared visibilities of both the target and
the calibrator stars following the method described in
\citet{Wittkowski01} and \citet{Hummel03}.

As we have discussed in \S~\ref{sec:setup}, $\gamma$~Cas could not be
observed on the AC--W7 baseline because of its high declination that
places the star outside the accessible sky coverage~(defined by the
limits imposed by the optical path compensation components).
$\phi$~Per was not constrained by the same limitation having slightly
lower declination.  Because the two stars are separated in right
ascension by less than 50~min it was more efficient to observe both
stars in the same mode~(i.e., using only 3 stations).  The loss of one
station for $\phi$~Per was compensated by the extended hour angle
range over which it was possible to observe this star~(recall
Fig.~\ref{fig:sky}).  Another advantage of having only one baseline on
each spectrograph is the elimination of multi-baseline cross-talk
effects that may be present in the data~\citep{Schmitt05}.  The
disadvantage of having only two simultaneous baselines is that some
interferometric observables such as closure phase and triple amplitude
could not be obtained because they require at least three simultaneous
baselines.

We observed $\gamma$~Cas and $\phi$~Per during November and December
of 2004.  Although we lost many nights due to poor weather, we still
managed to obtain $\sim$100 scans for each star with almost all the
scans providing data on two baselines.  Table~\ref{tab:obslog} shows a
log of the observations.  On 2004 Dec 8 the vacuum pipes between the
6th and 7th station on each of the three arms were disconnected~(see
Fig.~\ref{fig:baselines} for the layout of the imaging array) in order
to install safety vacuum valves.  This resulted in the W7 station
being unavailable for observations starting 2004 Dec 10, and
therefore, only observations utilizing the AE, AC and AW stations were
recorded for the rest of the observing run.  On 2004~Dec~2 we
temporarily removed the narrowband filters from the outputs of the
beam combiner to obtain observations without the use of the filters,
but with the same instrumental configuration.

For the purpose of the analysis presented in this paper, we are only
interested in the squared visibilities from the spectral channels that
contain the H$\alpha$ emission line.  However, before these $V^2$
values can be extracted from the observational data set, we need to
calibrate these quantities.  We follow the calibration procedure
outlined in \citet{Tycner03}, which we only summarize here.  The
calibration procedure uses the continuum channels at a given scan and
baseline to estimate a correction function that has a quadratic
dependence in $\lambda^{-1}$, which is then applied to the squared
visibility measure in the spectral channel that contains the H$\alpha$
emission line.  Because the correction function is a slowly varying
function across channels, it cannot account for any high order
channel-to-channel variations.  Therefore, in the current analysis we
estimated the $V^2$ amplitudes of the high order variations across the
spectral channels using the residuals that could not be removed with
the quadratic polynomials from the scans of the calibrator stars.  The
calibrators used for $\gamma$~Cas and $\phi$~Per were
$\epsilon$~Cas~(=HR~542) and $\zeta$~Cas~(=HR~153), respectively.
Both calibrators are B-type stars and were verified by spectroscopic
observations to have H$\alpha$ in absorption.  After confirming that
the channel-to-channel variations established based on observations of
the calibrators are stable on time scales longer than one night, we
calculated nightly averages that were then divided out of the scans of
the target stars.

\begin{figure}
\plotone{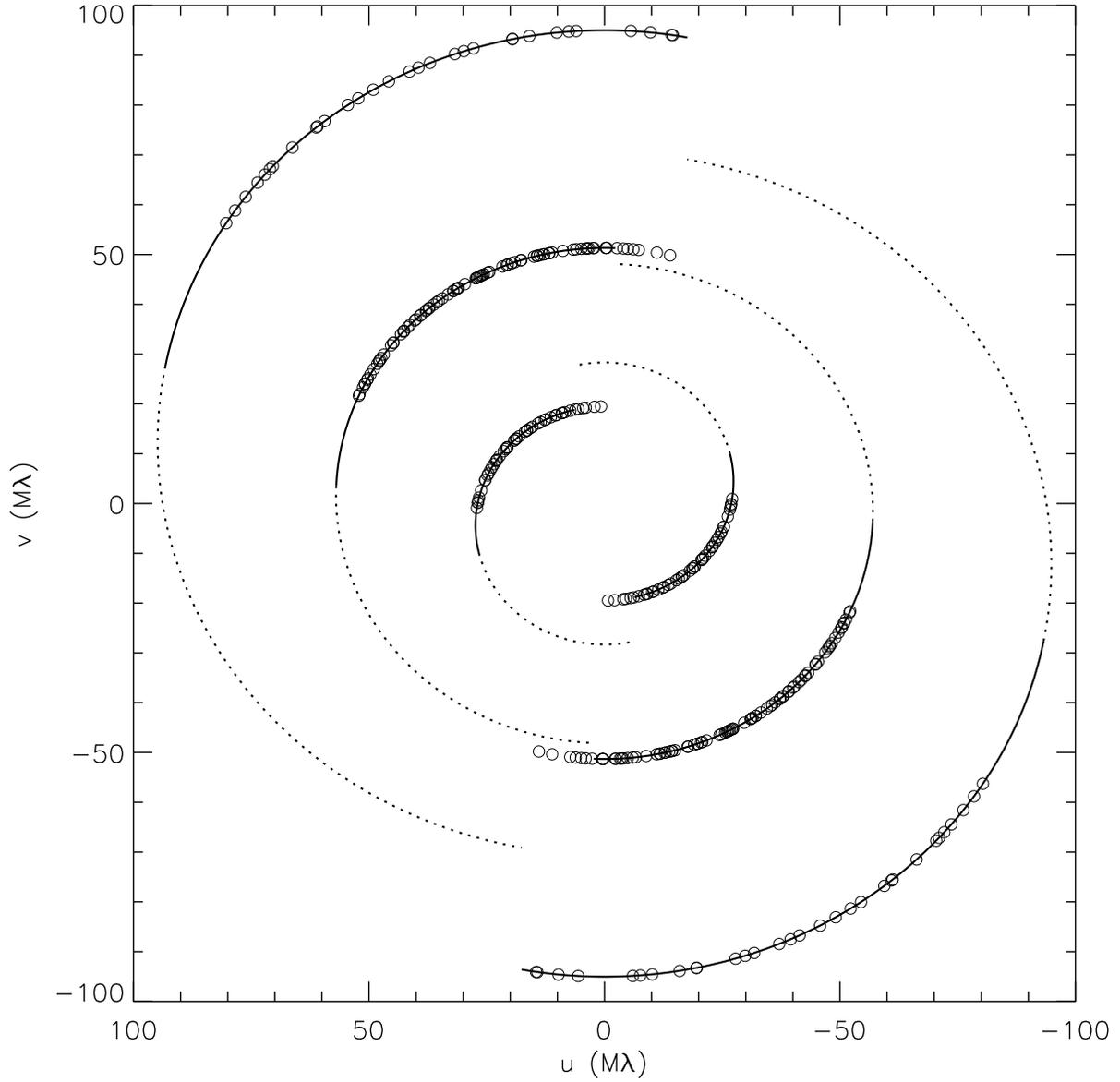}
\caption{Sampling of the ($u,v$)-plane by the H$\alpha$ observations
of $\gamma$~Cas on three baselines with lengths of 18.9~(AC--AE),
37.5~(AE--AW), and 64.4~m(AE-W7).  For comparison, sample
coverages~(not limited by the HA limitations shown in
Fig.~\ref{fig:sky}) over 6~hr ranges in HA from the meridian to the
east~({\it dotted line}) and to the west~({\it solid line}) are also
shown.}
\label{fig:gamCas-uv}
\end{figure}

\begin{figure}
\plotone{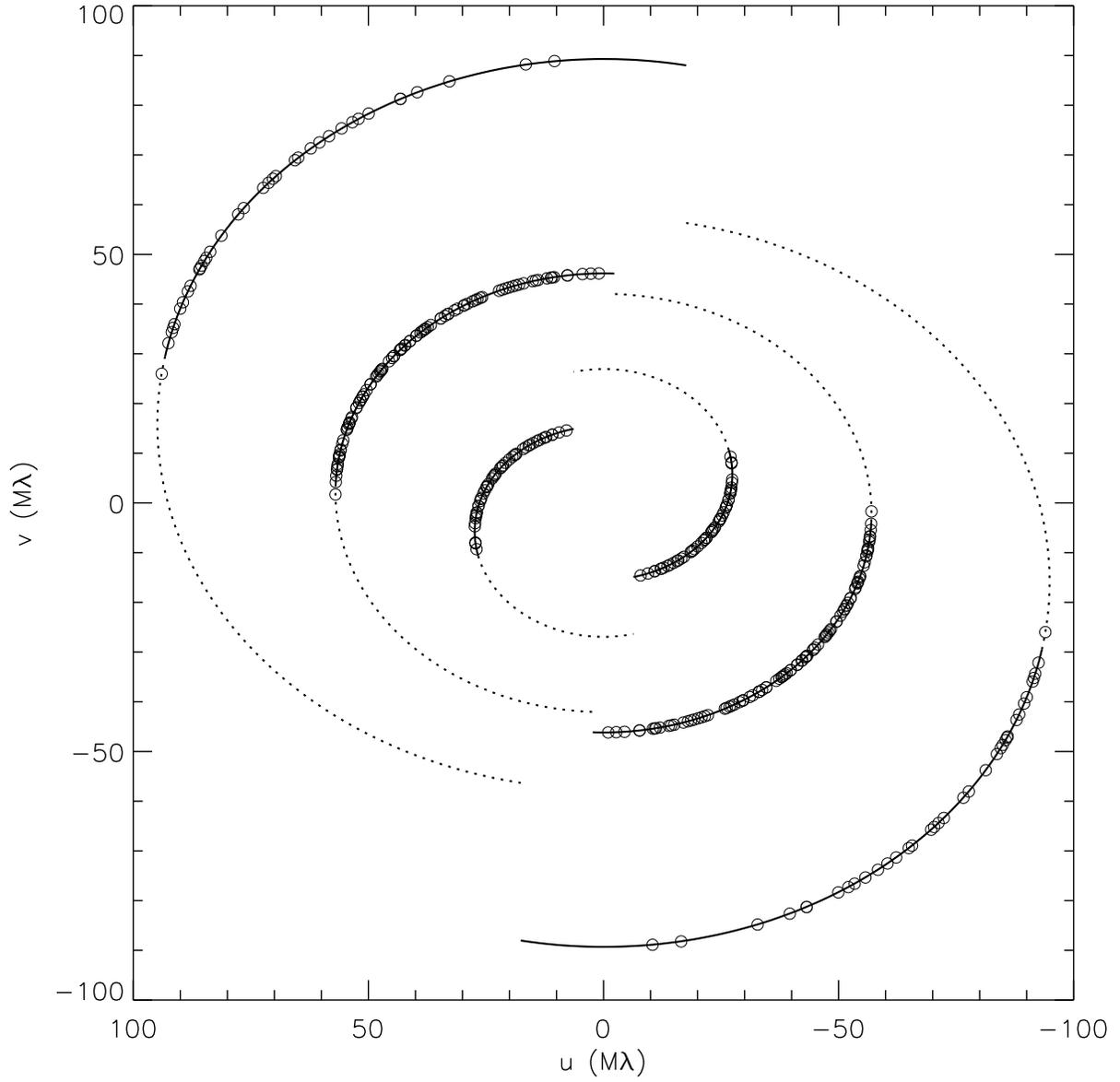}
\caption{Same as Fig.~\ref{fig:gamCas-uv} but for the ($u,v$)-plane
coverage of $\phi$~Per.  }
\label{fig:phiPer-uv}
\end{figure}

\begin{figure}
\plotone{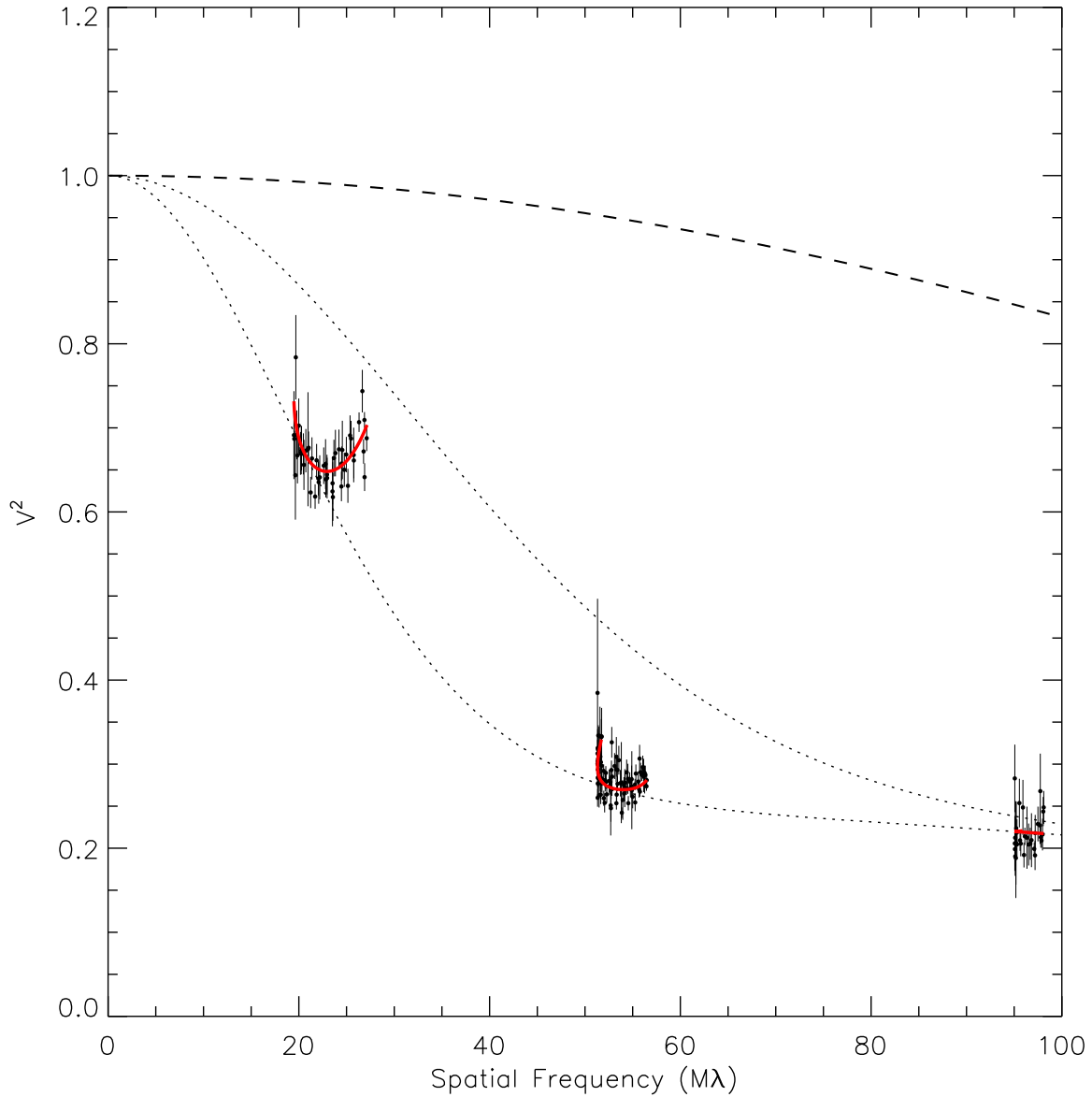}
\caption{Calibrated squared visibilities from the H$\alpha$ channel of
$\gamma$~Cas obtained at three baselines.  The elliptical Gaussian
model~({\it red solid line}) fitted to all observations is shown at
each baseline for the same HA range as defined by the observations.
The model contains a contribution from the stellar photospheric
disk~({\it dashed line}) that is modeled with
equation~(\ref{eqn:star}).  Model curves evaluated at the minor~({\it
upper dotted line}) and major~({\it lower dotted line}) axes of the
elliptical Gaussian model are also shown.}
\label{fig:gamCas-all-data-fit}
\end{figure}

\begin{figure}
\plotone{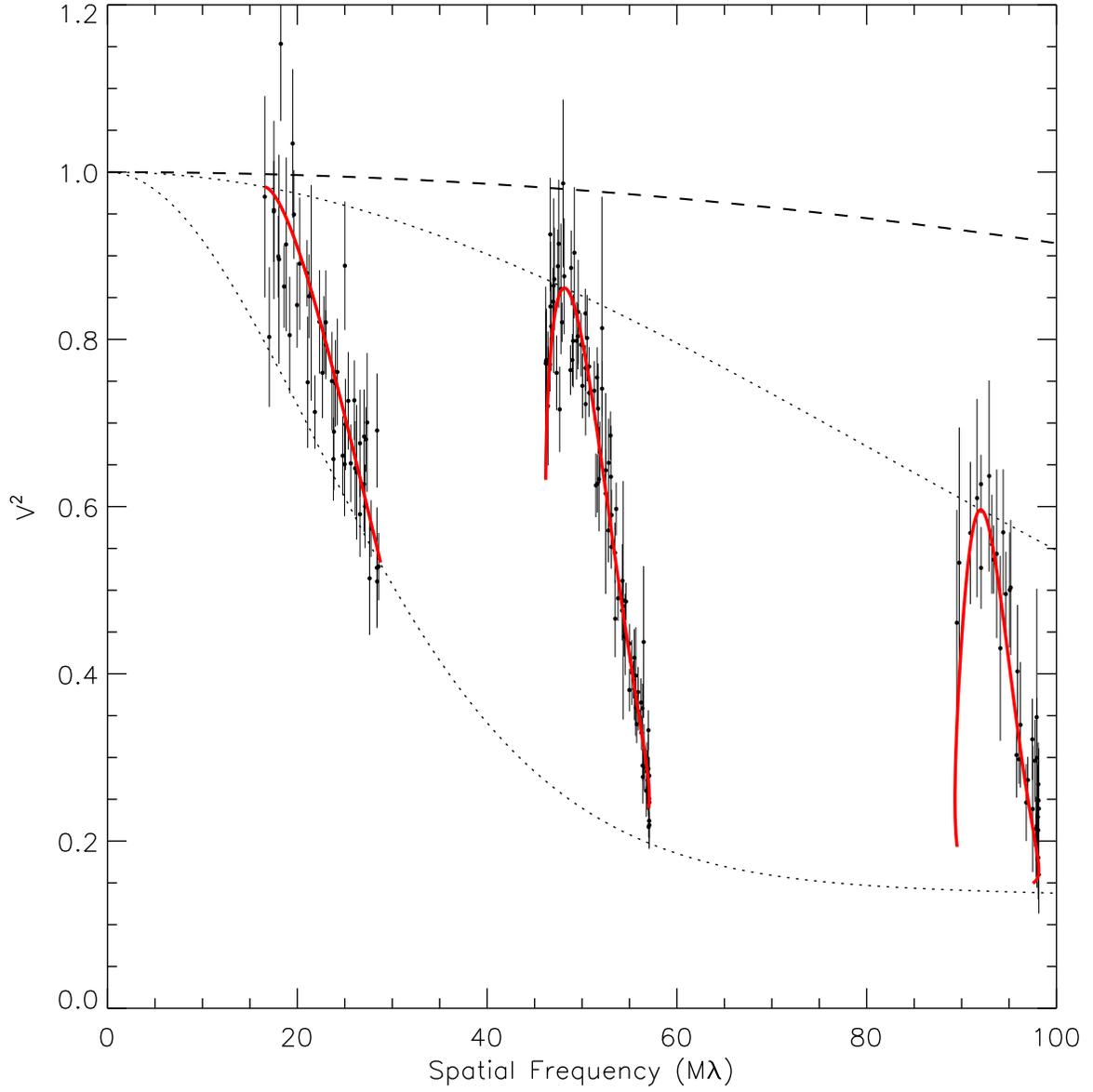}
\caption{Same as Fig.~\ref{fig:gamCas-all-data-fit} but for the
H$\alpha$ observations of $\phi$~Per.}
\label{fig:phiPer-all-data-fit}
\end{figure}

\begin{figure}
\plotone{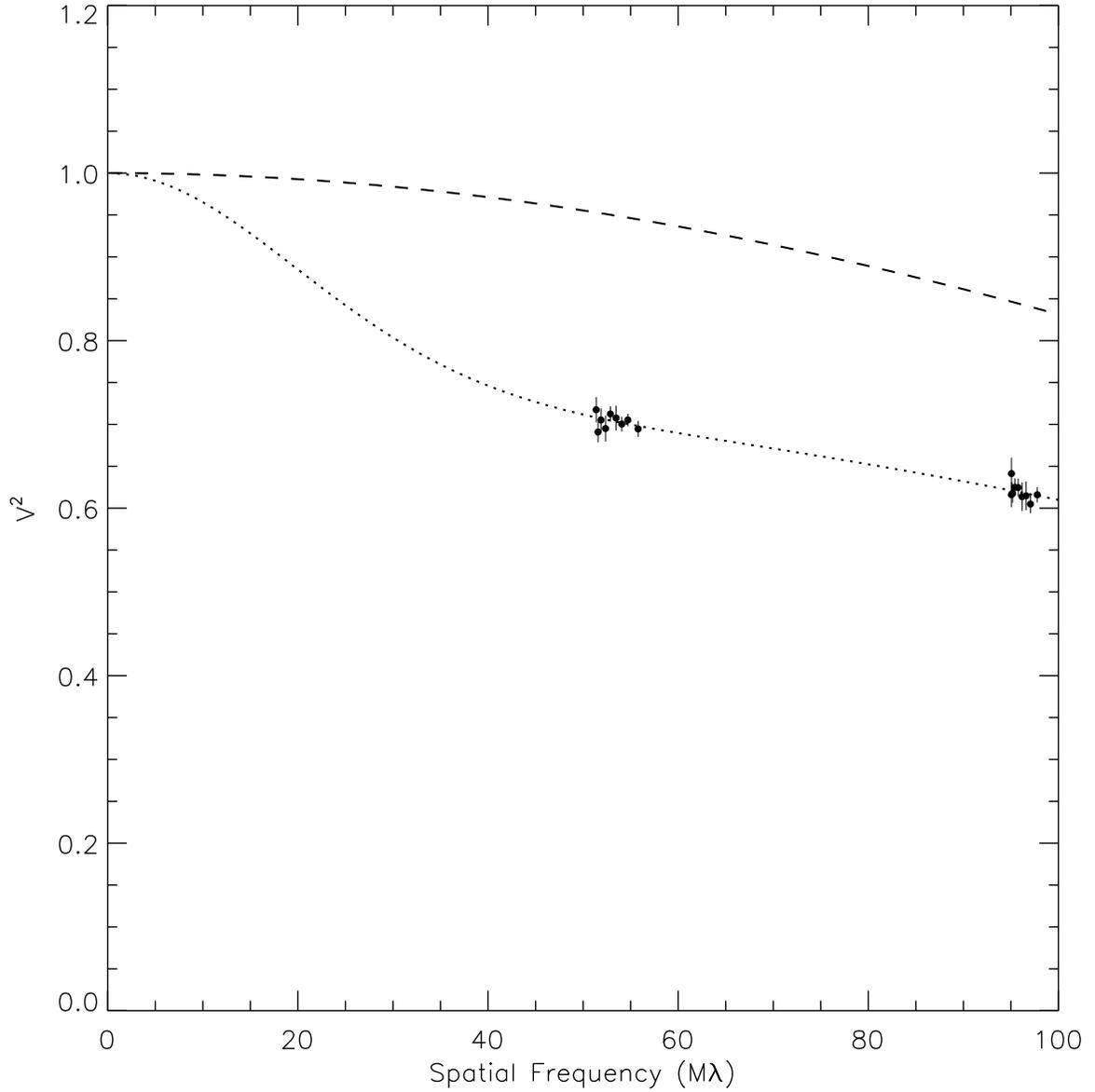}
\caption{Calibrated squared visibilities from the H$\alpha$ channel of
$\gamma$~Cas obtained on 2004~Dec~2 without the use of narrowband
filter at two baselines~(AE--AW and AE--W7).  The best-fit circularly
symmetric Gaussian model~({\it dotted line}) and the stellar
photospheric disk~({\it dashed line}) are also shown.
}
\label{fig:gamCas-no-filter}
\end{figure}

\begin{figure}
\plotone{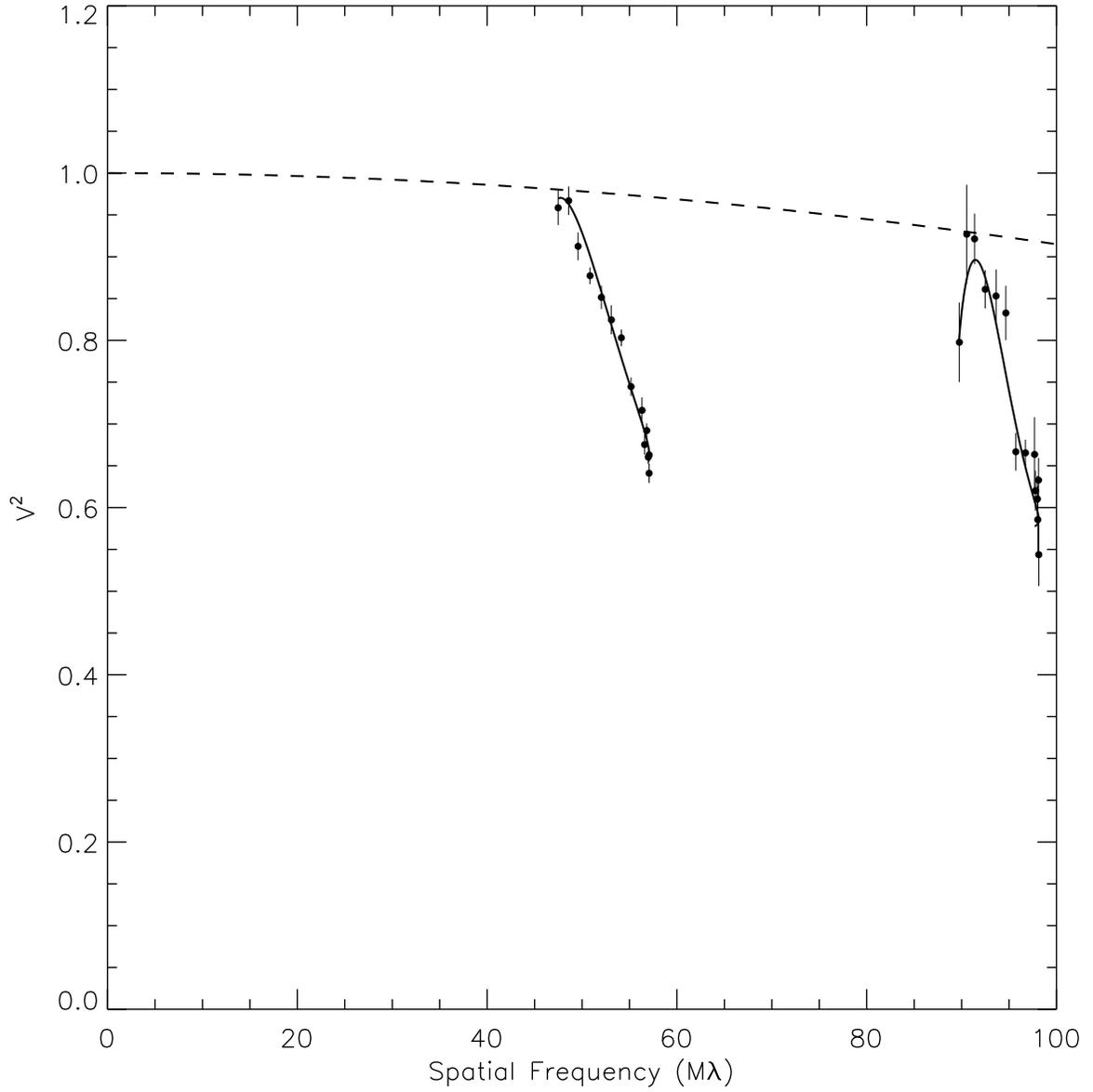}
\caption{Same as Fig.~\ref{fig:gamCas-no-filter} but for the H$\alpha$
observations of $\phi$~Per with the best-fit elliptical Gaussian model
shown~({\it solid lines}).}
\label{fig:phiPer-no-filter}
\end{figure}

After applying the bias and the calibration corrections to all scans,
we extract the squared visibilities for the H$\alpha$ channel forming
two large data sets, one for $\gamma$~Cas~(with 169 data points) and
the other for $\phi$~Per~(with 186 data points).  The observations
obtained on 2004~Dec~2, when the narrowband filters were not used, are
treated as separate data sets.  Figures~\ref{fig:gamCas-uv}
and~\ref{fig:phiPer-uv} show the resulting ($u,v$)-plane coverage at
the AC--AE, AE--AW, and AE--W7 baselines in the H$\alpha$ channel for
$\gamma$~Cas and $\phi$~Per, respectively.  The corresponding $V^2$
data for the two stars are shown in
Figures~\ref{fig:gamCas-all-data-fit}
and~\ref{fig:phiPer-all-data-fit} for observations obtained with the
narrowband filter, whereas the $V^2$ observations obtained without the
filters are shown in Figures~\ref{fig:gamCas-no-filter} and
\ref{fig:phiPer-no-filter}.

\subsection{Spectroscopy}
\label{sec:spectroscopy}

\begin{figure}
\plotone{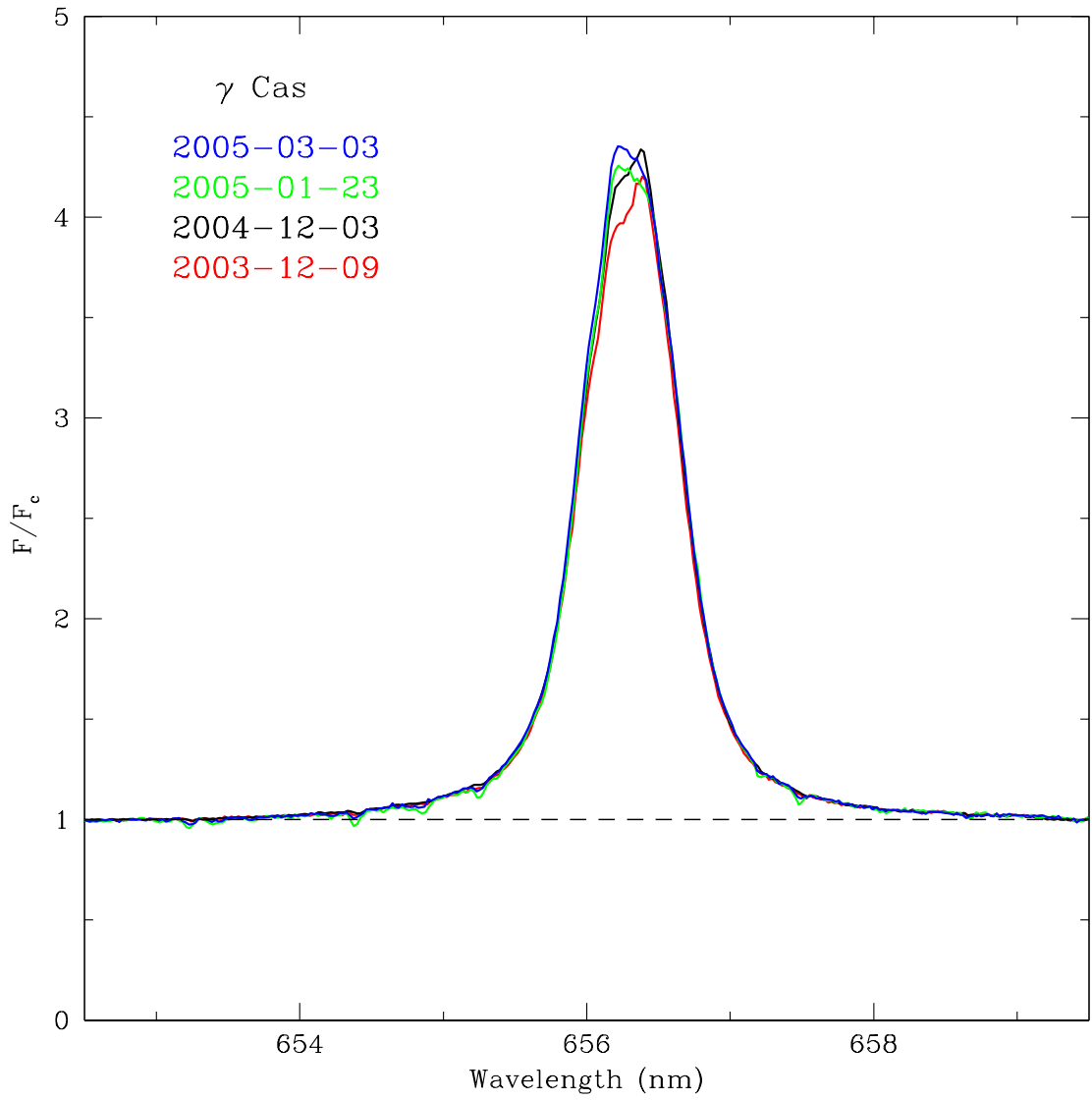}
\caption{H$\alpha$ profiles of $\gamma$~Cas obtained at four different
epochs.  }
\label{fig:Halpha_GCas}
\end{figure}

\begin{figure}
\plotone{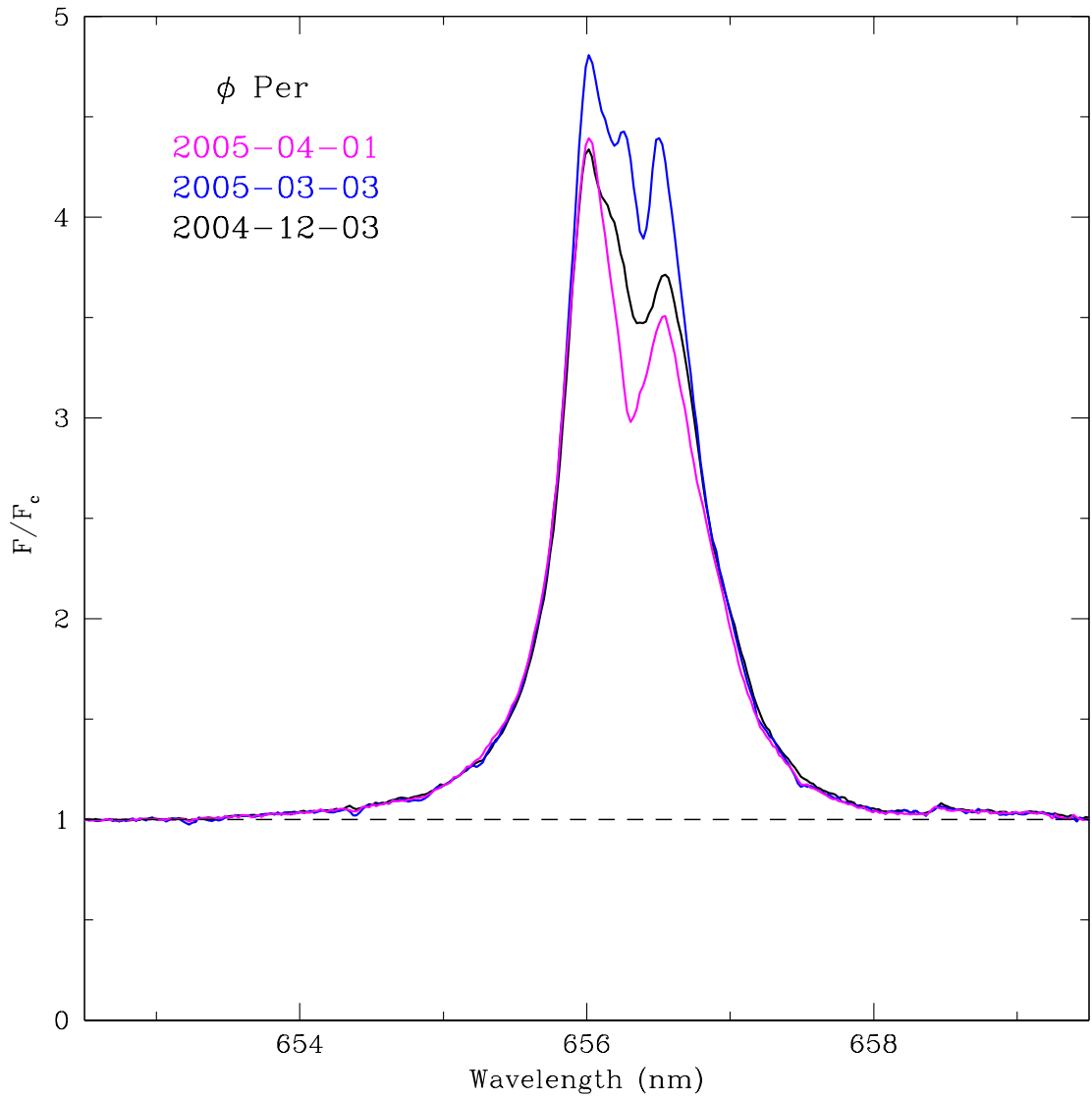}
\caption{H$\alpha$ profiles of $\phi$~Per obtained at three different
epochs.  }
\label{fig:Halpha_PPer}
\end{figure}

The interferometric observations obtained with the NPOI do not contain
enough spectral information to help us establish the properties of the
H$\alpha$ emission line.  Therefore, we observed $\gamma$~Cas and
$\phi$~Per using a fiber-fed echelle spectrograph~(also known as the
Solar-Stellar Spectrograph; SSS), which is located at the Lowell
Observatory's John S.~Hall 1.1~m telescope.  The SSS instrument
produces spectra in the H$\alpha$ region with the resolving power of
$\sim$10,000 and a signal-to-noise ratio~(SNR) of few hundred in the
H$\alpha$ region.  The SSS observations were reduced using standard
reduction routines written in IDL\footnote{Interactive Data Language
of RSI, ITT Industries, Boulder, CO}~\citep{Hall94}, and the resulting
H$\alpha$ profiles of $\gamma$~Cas and $\phi$~Per are shown in
Figures~\ref{fig:Halpha_GCas} and~\ref{fig:Halpha_PPer}, respectively.

The H$\alpha$ emission of $\gamma$~Cas is known to be stable on time
scales greater than $1$~yr and our spectra confirm this.  The EWs of
the four H$\alpha$ profiles shown in Figure~\ref{fig:Halpha_GCas}
range from $-3.0$ to $-3.2$~nm.  Because the largest uncertainty
associated with these values is the precision of continuum
normalization, which we estimate at 3\%, for the purpose of our study
we use an EW of $-3.1\pm 0.1$~nm for the H$\alpha$ emission of
$\gamma$~Cas.

As we will discuss in \S~\ref{sec:disk_truncation}, $\phi$~Per is a
Be+sdO spectroscopic binary with a period of 127~d and is known to
show H$\alpha$ variability.  The observed variability is attributed in
part to the radiative effects of the hot secondary on the
circumstellar region associated with the primary.  Although our
temporal coverage is limited, we confirm the presence of H$\alpha$
emission variability in $\phi$~Per.  The equivalent widths of the
three profiles shown in Figure~\ref{fig:Halpha_PPer} are $-3.8$,
$-4.2$, and $-3.6$~nm, which were obtained on 2004~Dec~3, 2005~Mar~3,
and 2005~Apr~1, respectively.  However, because the relative shape of
the H$\alpha$ profile does not appear to change significantly, we
conclude that combining all of the interferometric observations from
our observing run should not result in any significant errors~(we will
return to this point in \S~\ref{sec:model_fitting}).  For this reason
we adopt an H$\alpha$ EW for $\phi$~Per of $-3.9\pm 0.3$~nm, where the
larger uncertainty should account for the intrinsic variability during
our interferometric run.

\section{The Analysis}

\subsection{Models}
\label{sec:models}

The squared visibilities from the spectral channels containing the
H$\alpha$ emission contain two signatures, one due to the central star
and the other due to the circumstellar region.  Therefore, these data
must be modeled with a two component model of the form
\begin{equation}
\label{eq:model}
V^2_{\rm model} = \left[ c_p V_p + (1-c_p)V_{\rm env} \right]^2,
\end{equation}
where $V_p$ and $V_{\rm env}$ are the visibility functions
representing the photosphere of the central star and the circumstellar
envelope, respectively, and $c_p$ is the fractional contribution from
the stellar photosphere to the total flux in the H$\alpha$ channel.
Generally, both $V_p$ and $V_{\rm env}$ can be complex functions, but
because the models we consider in this analysis are all
point-symmetric, as well as concentric, the functions are real and can
be treated as visibility amplitudes.

To decrease the number of free parameters in our models we treat the
angular diameter of the central star as a known quantity.  This is
also supported by the fact that the central stars are not
significantly resolved even at our longest baseline.  Furthermore, as
we discussed in \citet{Tycner04}, our best-fit models describing the
H$\alpha$-emitting regions are not very sensitive to the assumed
diameter of the central star~(changing the diameter by a factor of 2
results in $\sim$~1~\% change in the best-fit disk parameters).  For
the same reason we can also ignore any effects related to the rapid
rotation for which Be stars are well known for, because the
geometrical distortion and gravity darkening near the stellar equator
will only weakly affect our best-fit disk parameters.

The most widely used approach for predicting a stellar angular
diameter is to use empirically established relationships between the
stellar color index and the surface brightness, which in turn is
related to the angular size~\citep{Barnes78,vanBelle99}.  Another
approach would be to use tabulations of linear stellar diameters as a
function of spectral class~\citep{Underhill79}.  However, because
there is an intrinsic scatter in stellar diameters at each spectral
class, and the distance to the source needs to be known to convert the
linear diameter to the angular size needed for interferometry, this is
not the preferred approach.  Therefore, for $\gamma$~Cas and
$\phi$~Per we adopt the same stellar angular diameters as were used by
\citet{Quirrenbach97}, which were derived using the photometric
relations of \citet{Barnes78}.  The angular diameters we adopt for
$\gamma$~Cas and $\phi$~Per are 0.56 and 0.39~mas, respectively.

We model the stellar photosphere of each star with a circularly
symmetric uniform disk brightness distribution.  This ensures that the
central star is modeled in exactly the same way, at the continuum
channels during the calibration procedure~\citep[see][for more
details]{Tycner03} and at the H$\alpha$ channel via
equation~(\ref{eq:model}).  Therefore, we represent the visibility
function of the photospheric component of an angular diameter $a$ with
\begin{equation}
\label{eqn:star}
V_{p}(u,v) = 2 \frac{J_1(\pi a \sqrt{u^2 + v^2})}{\pi a \sqrt{u^2 +
v^2}},
\end{equation}
where $J_1$ is a first-order Bessel function and $u$ and $v$ are the
spatial frequencies, which are given by the east-west and south-north
components of the projected baseline on the plane of the sky divided
by wavelength~\citep[see \S4.1 in][for the definition of $u$-$v$
plane]{Thompson01}.

Our interferometric observations were obtained at high spatial
frequencies~(i.e., at high spatial resolution), and therefore, we can
compare different models of brightness distribution.  Previously
published work on disk models of Be stars has demonstrated that the
thermal structure of the circumstellar disks can be quite
complex~\citep{Millar98,Millar00,Carciofi06}.  However, for this first
investigation with this new observational technique we choose to model
our data with three simple models, uniform disk~(UD), uniform
ring~(UR), and a Gaussian distribution~(GD).  The visibility
amplitudes for all three models can be written as
\begin{equation}
\label{eqn:envelope}
V_{\rm env} = \left\{
\begin{array}{c@{\quad:\quad}l}
2 J_1(\pi \theta_{\rm UD} s) /\pi \theta_{\rm UD} s  &  {\rm UD}\\
2 (1-\epsilon^2)^{-1} [J_1(\pi \theta_{\rm UR} s) /\pi \theta_{\rm UR} s -
\epsilon^2 J_1(\pi \epsilon \theta_{\rm UR} s) /\pi \epsilon \theta_{\rm UR} s] &
{\rm UR}\\
\exp [- (\pi \theta_{\rm GD} s)^2 / 4\ln 2]      & {\rm GD}
\end{array} \right. ,
\end{equation}
where $\theta_{\rm UD}$, $\theta_{\rm UR}$ correspond to the angular
diameters of the UD and UR models, respectively, and $\theta_{\rm GD}$
corresponds to the full-width at half-maximum~(FWHM) of the Gaussian
model.  Because we allow all three models to have elliptical
distribution on the sky, all of the above diameters describe the
dimensions along the major axis.  The other variables in
equation~(\ref{eqn:envelope}) are, $\epsilon$ that describes the inner
diameter of the ring model along the major axis~(in units of
$\theta_{\rm UR}$), and $s$ that is given by
\begin{equation}
\label{eqn:s}
s = \sqrt{r^2 (u \cos \phi - v \sin \phi)^2 + (u \sin \phi + v \cos
\phi)^2},
\end{equation}
where $r$ is the axial ratio and $\phi$ is the position angle~(PA;
measured east from north) of the major axis~(when $0 \leq r < 1$).
For circularly symmetric structures~($r=1$), equation~(\ref{eqn:s})
reduces to a simple expression for a radial spatial frequency in the
$(u,v)$-plane that is given by $(u^2+v^2)^{1/2}$.

\subsection{Model Fitting}
\label{sec:model_fitting}

To obtain a best-fit for each of the three models, we used a nonlinear
least-squares IDL procedure based on the Levenberg-Marquardt
method~\citep{Press92}.  Each model is represented by
equation~(\ref{eq:model}), with the appropriate expression for the
envelope component from equation~(\ref{eqn:envelope}), and a fixed
visibility function for the photospheric
component~(eq.~[\ref{eqn:star}] with a fixed stellar diameter $a$).
Therefore, the UD and GD models have four free parameters, $c_p$,
$\theta$, $r$, and $\phi$, whereas the UR model has $\epsilon$ as an
extra parameter.  Table~\ref{tab:best_fit} lists the best-fit values
for the parameters for all three models along with the reduced
$\chi^2$ values.  By inspecting the reduced $\chi^2$ values in
Table~\ref{tab:best_fit} for the three $\gamma$~Cas models, it is
evident that the UD and UR models cannot represent the observations
obtained at all three baselines.  Therefore, we have also fitted
models to observations of $\gamma$~Cas from the shortest two baselines
only, where all three models yield similar quality fits~(these models
are also listed in Table~\ref{tab:best_fit}).  To check for any
possible variations in the models that could be attributed to the
temporal variability of our sources~(especially in the case of
$\phi$~Per), we fitted models to various subsets of our large data
set.  Although typically the uncertainties of the best-fit parameters
are larger for models fitted to smaller data sets, all results were
consistent with our solutions based on all observations.

\begin{figure}
\plotone{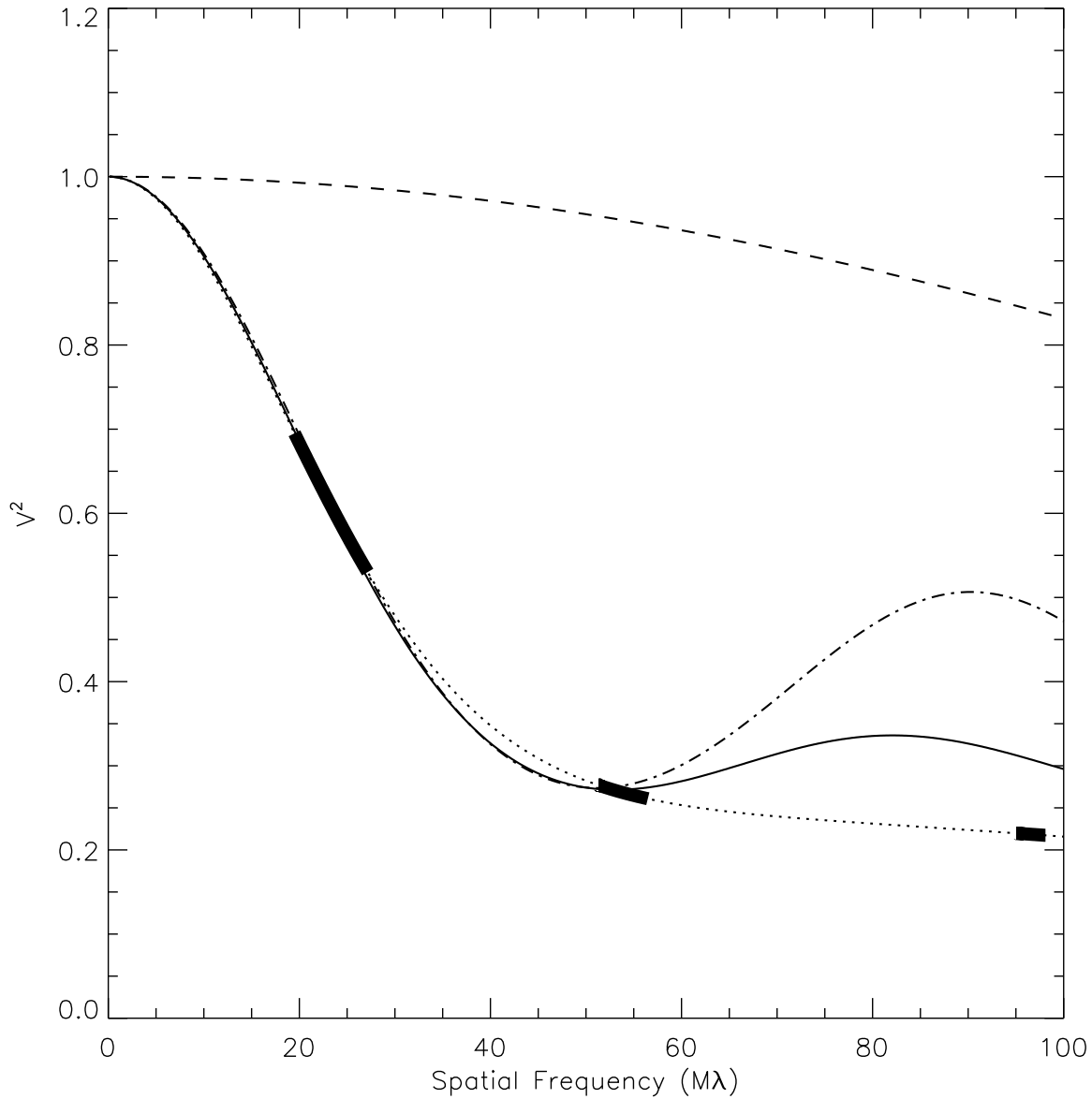}
\caption{Model squared visibilities for the uniform disk~({\it solid
line}), ring~({\it dash-dot line}), and Gaussian~({\it dotted line})
models of $\gamma$~Cas.  All three models contain a contribution from
a central star~({\it dashed line}), which is modeled using
eq.~(\ref{eqn:star}), and are evaluated along their major axes~(i.e.,
along their largest angular extent).  The spatial frequency ranges
sampled by the three baselines are indicated with thick solid lines.
The models shown were fitted to the data at the two shortest baselines
only. }
\label{fig:compare_modelsGC}
\end{figure}

To illustrate why the UD and UR models fail to represent the
observations of $\gamma$~Cas obtained at all three baselines, in
Figure~\ref{fig:compare_modelsGC} we plot the best-fit UD, UR, and GD
models, which were only fitted to data from the shortest two
baselines.  Each baseline provides data over a different spatial
frequency range~(because of different physical lengths) and these
ranges are marked with thick solid lines in
Figure~\ref{fig:compare_modelsGC}.  A small fraction of each range
corresponds to an instant at which the projected baseline on the sky
was oriented along the major axis for which these model curves were
evaluated.  The model curves demonstrate the inherent degeneracy
present between the three models at low spatial frequencies~(i.e., at
low spatial resolutions the three models look the same).  However, the
longest baseline provides data at high enough spatial frequencies so
that the degeneracy between the three models is eliminated~(see also
Fig.~\ref{fig:gamCas-all-data-fit}).  We conclude that the UD and UR
models are inconsistent with the data obtained at the longest
baseline~(largest spatial frequencies), whereas the GD models fit the
observations at all baselines.

\begin{figure}
\plotone{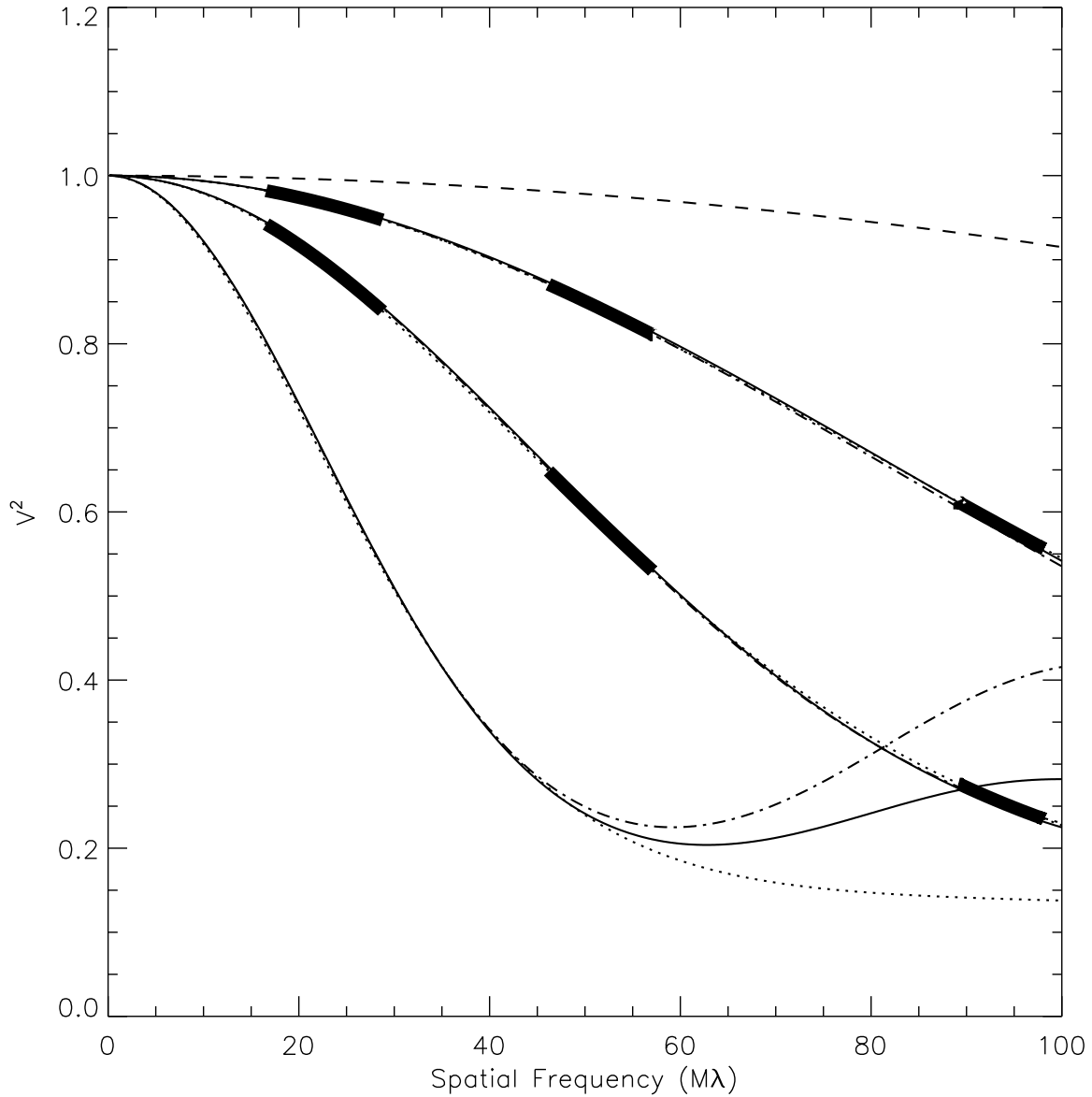}
\caption{Same as Fig.~\ref{fig:compare_modelsGC} but for models based
on observations of $\phi$~Per and evaluated at three different
orientations with respect to the major axis.  The three families of
curves correspond to 100\%~({\it lowest three curves}), 50\%~({\it
three overlapping curves in the center}), and 27\%~({\it top three
overlapping curves}) of the size of the major axis.  Because the
observations do not sample the disk along the major axis, the spatial
frequency ranges sampled by the three baselines are indicated only
along the minor and half-major axes~({\it thick solid lines}).}
\label{fig:compare_modelsPP}
\end{figure}

The observations of $\phi$~Per are more challenging because the
angular size of its circumstellar region is smaller than for
$\gamma$~Cas, as well as the projected baselines on the plane of the
sky are not oriented along its major axis.  This results in
interferometric observations that do not resolve the region
sufficiently well to break the degeneracy between the different
models.  This is also the reason all three models for $\phi$~Per
listed in Table~\ref{tab:best_fit} yield the same $\chi^2_{\nu}$
values.  Figure~\ref{fig:compare_modelsPP} demonstrates the
differences between the model curves based on the three best-fit
models evaluated at three different projections.  The three
projections were chosen to correspond to the minor axis~(that is 27\%
of the major axis), along a direction where the extent is 50\% of the
major axis, and along the major axis.  Because we do not have
observations that resolve the major axis~(recall
Fig.~\ref{fig:phiPer-all-data-fit}), the spatial frequency ranges over
which we have data are marked only on the curves that correspond to
the minor axis and 50\% of the major~(marked with thick solid lines in
Fig.~\ref{fig:compare_modelsPP}).  Based on
Figure~\ref{fig:compare_modelsPP} we conclude that in order to break
the degeneracy between the models, either observations at higher
spatial frequencies, or along the major axis, are required~(as
demonstrated in the case of $\gamma$~Cas).

\subsection{Gaussian Distribution}
\label{sec:gauss}

For $\phi$~Per we cannot exclude the UD and UR models in favor of the
GD model based on interferometric data alone, however, we can
demonstrate that the GD model is the preferred solution.  This is
because the three models that fit the $\phi$~Per data equally well,
have different best-fit values for the $c_p$ parameter, which can be
constrained by spectroscopy.  For example, if the EW of the continuum
flux from the star in the H$\alpha$ channel is $F_{\star}$~(which has
units of length) and the EW of the net H$\alpha$ emission is $E_{{\rm
H}\alpha}$, then the fractional photospheric contribution in the
absence of a narrowband filter~($c_p^{\star}$) can be expressed as
\begin{equation}
\label{eqn:cp_star}
c_p^{\star} = \frac{F_{\star}}{F_{\star} + E_{{\rm H}\alpha}}.
\end{equation}
With the narrowband filter $F_{\star}$ gets reduced to $t F_{\star}$,
where $t$ is the fractional transmission in the H$\alpha$ channel.
The H$\alpha$ emission gets reduced by the filter as well, but because
the H$\alpha$ emission lines have FWHM $\lesssim$ 1nm, which is less
than the width of 2.8~nm of the narrowband region, we can approximate
the transmission at the H$\alpha$ line~($\eta$) by the peak
transmission of 92~\% of the narrowband region~(recall
Fig.~\ref{fig:filter}).  Therefore, the $c_p$ parameter for the case
including the filter can be written as
\begin{equation}
\label{eqn:cp_filter}
c_p = \frac{t F_{\star}}{t F_{\star} + \eta E_{{\rm H}\alpha}}.
\end{equation}
Taking the ratio of equations~(\ref{eqn:cp_star})
and~(\ref{eqn:cp_filter}) allows us to obtain an expression for $t$ of
the form
\begin{equation}
\label{eqn:t}
t = \eta \frac{1/c_p^{\star} -1 }{1/c_p -1},
\end{equation}
which does not depend on $F_{\star}$ or $E_{{\rm H}\alpha}$.  

Equation~(\ref{eqn:t}) requires the photospheric contributions $c_p$
and $c_p^{\star}$ to be known.  For $c_p$ we use the values listed in
Table~\ref{tab:best_fit} obtained for the GD model fits to
observations with the narrowband filters.  To obtain the corresponding
values for $c_p^{\star}$, we fit GD models to observations obtained
without the filters (i.e., the observations obtained on 2004~Dec~2).
The squared visibilities for the best-fit models to the 2004~Dec~2
data are shown in Figures~\ref{fig:gamCas-no-filter}
and~\ref{fig:phiPer-no-filter}, for $\gamma$~Cas and $\phi$~Per,
respectively.  In both cases the best-fit GD model parameters~(except
for $c_p$) agree with the parameters listed in
Table~\ref{tab:best_fit}, even for $\gamma$~Cas where we did not have
enough points to constrain an elliptical model.  These fits yield
$c_p^{\star}$ values of 0.85 and 0.78 for $\gamma$~Cas and $\phi$~Per,
respectively.  Using equation~(\ref{eqn:t}) with the appropriate
values for $\gamma$~Cas and $\phi$~Per we obtain the same estimate for
$t$ of $0.17 \pm 0.01$.  It is interesting to note that this 17~\%
transmission is fully consistent with the detected drop in photon
counts in the H$\alpha$ channel when the narrowband filters are
introduced for observations of calibrator stars~(which do not have
H$\alpha$ in emission).

Because our estimate for $t$ does not dependent on $F_{\star}$ or
$E_{{\rm H}\alpha}$, we can use equation~(\ref{eqn:cp_filter}) to
obtain a value for $c_p$ based on spectroscopic observations.  We
approximate $F_{\star}$ with the spectral width of the H$\alpha$
channel~(which has been measured to be $15 \pm 1$~nm).  To obtain the
EW of the net H$\alpha$ emission for $\phi$~Per, we use the H$\alpha$
EW from \S~\ref{sec:spectroscopy} and we add a small
contribution~(0.37~nm) due to the absorption line that has been filled
in~\citep[using the same procedure as outlined in \S~4.2
of][]{Tycner05}.  The net H$\alpha$ emission of $4.27 \pm 0.30$~nm for
$\phi$~Per results in $c_p$ of $0.39 \pm 0.03$.  This estimate is
significantly lower than the values listed in Table~\ref{tab:best_fit}
for the UD and UR models, which have best-fit $c_p$ values of 0.52 and
0.62, respectively, but it is fully consistent with the $0.39\pm 0.02$
value obtained for the GD model.  We interpret this as an indication
that the brightness distribution of the circumstellar disk of
$\phi$~Per is best represented by a GD model.  It is interesting to
note, if we follow the same reasoning for $\gamma$~Cas data from the
shortest two baselines~(where the degeneracy between the three models
also exists) we arrive at the same conclusion that the GD model is the
preferred solution.  For $\gamma$~Cas, based on its H$\alpha$ EW of
$-3.1 \pm 0.1$~nm and an estimated component accounting for the
absorption line of 0.26~nm, we estimate that $c_p = 0.45 \pm 0.03$,
which again is closest to the value obtained for the GD model.

We conclude that out of the three models we have considered in our
analysis the GD model is the only model that fits $\gamma$~Cas
observations and is the preferred model for $\phi$~Per based on
spectroscopic constraints.  Therefore, in Table~\ref{tab:parameters}
we list our final best-fit parameters describing the circumstellar
regions of $\gamma$~Cas and $\phi$~Per based on the GD model fits.
The model squared visibilities calculated based on these models are
shown with red solid lines in Figures~\ref{fig:gamCas-all-data-fit}
and~\ref{fig:phiPer-all-data-fit} for $\gamma$~Cas and $\phi$~Per,
respectively.

\section{Discussion}

\subsection{Circumstellar Region}

Because the number of published parameters describing the
circumstellar regions of $\gamma$~Cas and $\phi$~Per is still very
small, it is worthwhile to compare our results to those published in
the literature.  Both, \citet{Quirrenbach97} and \citet{Tycner03}
fitted a GD model to their observations and therefore their model
values can be compared directly with those listed in
Table~\ref{tab:parameters}.

Using observational data from the Mark~III interferometer
\citet{Quirrenbach97} obtained best-fit model parameters for
$\gamma$~Cas of $\theta = 3.47 \pm 0.02$~mas, $r=0.70\pm0.02$, and PA
of 19$\pm2^{\circ}$.  \citet{Tycner03} reported results based on older
NPOI observations~(with a maximum baseline of 37.5~m), which had
$\theta = 3.67 \pm 0.09$~mas, $r=0.79\pm0.03$, and PA of
32$\pm5^{\circ}$.  Our new determination of the angular size of the
major axis of $3.59 \pm 0.04$, fully confirms the previous estimates.
The apparent axial ratio of $0.58 \pm 0.03$ that we find for
$\gamma$~Cas is smaller than the published values and this might be
related to the fact that our interferometric observations cover much
larger range of baseline projection angles on the plane of the
sky~(recall Fig.~\ref{fig:gamCas-uv}).  Our best-fit value for the PA
of $31\fdg2 \pm 1\fdg2$ agrees with our previous
estimate~\citep{Tycner03}, but differs from the PA reported by
\citet{Quirrenbach97} whose value was at right angle to the
polarization vector they obtained from polarimetry.  Possible sources
for this discrepancy could be the intrinsic variability of the
polarization of the source~\citep[see for example Table~6
in][]{Quirrenbach97}, the residual effects associated with the removal
of the interstellar polarization, or the effects of non-axisymmetric
scattering surface~(i.e., only for axisymmetric sources can one expect
the polarization vector to be \emph{exactly} perpendicular to the
plane of the disk).

For $\phi$~Per, \citet{Quirrenbach97} obtained model values of $\theta
= 2.67 \pm 0.20$, $r=0.46\pm0.04$, and PA of $-62\pm5^{\circ}$, but
they point out that the value for $r$ might be an upper limit because
their baselines were not long enough to resolve the minor axis.
Indeed, our results based on observations that utilize baselines that
are more than twice as long than those used in their study yield a
best-fit value of $0.27\pm 0.01$ for the axial ratio.  However, the
angular size of the major axis of $2.89 \pm 0.09$ and PA of $-61\fdg5
\pm 0\fdg6$ that we obtain in this study fully agree with their
values.  Since $\phi$~Per was not observed by any other interferometer
than Mark~III, our results represent the first confirmation of the
values reported by \citet{Quirrenbach97}.

\subsection{Inclination and Disk Opening Angles}

Our interferometric observations of the spatially resolved
circumstellar regions, which show an apparent ellipticity, can be used
to estimate the inclination\footnote{The angle between the normal to
the plane of the disk and the line-of-sight.} and disk-opening angles.
For example, if we assume that the circumstellar disk is circularly
symmetric, we can obtain a lower limit on the inclination angle~($i$)
using the observed axial ratio~($r$) using
\begin{equation}
i \gtrsim \cos^{-1} r.  
\end{equation}
The minimum value corresponds to a geometrically thin disk, where the
entire signature of the apparent ellipticity is interpreted as a
projection effect in this case.  Similarly, if we assume that the
geometry of the circumstellar region can be represented with a simple
equatorial disk model with a half-opening angle $H_{\theta}$ and
radius $R_{\rm disk}$~\citep[see for example Fig.~2 in][]{Waters87},
we can then obtain an upper limit on $H_{\theta}$ using
\begin{equation}
H_{\theta} \lesssim \sin^{-1}(r/2),
\end{equation}
where the maximum value corresponds to a system viewed edge-on.

In Table~\ref{tab:parameters} we list the estimated limits on the
inclination and half-opening angles for both $\gamma$~Cas and
$\phi$~Per based on their best-fit values of $r$ obtained from the
elliptical Gaussian models.  Interestingly, for $\phi$~Per the lower
limit of $\approx$~74$^{\circ}$ for $i$ is consistent with the
80$^{\circ}$ to $88^{\circ}$ range originally proposed by
\citet{Poeckert81}.  Although \citet{Poeckert81} did not describe the
circumstellar region in terms of an opening angle, the disk thickness
of 1.3--2.7~$R_{\star}$ and the disk radius of 7.7~$R_{\star}$ can be
translated to a half-opening angle between 5 and 10$^{\circ}$, which
again is consistent with our upper limit of $\approx 8^{\circ}$.

It is instructive to compare the upper limits we obtain for
$H_{\theta}$ of $\gamma$~Cas and $\phi$~Per to the value of
15$^{\circ}$ adopted by \citet{Waters87} in their study of almost 60
Be stars based on far-IR IRAS observations.  \citet{Waters87}
estimated a number of different Be stellar characteristics, such as
the mass loss rate and the disk density, which had a dependence on
$H_{\theta}$.  Although the authors assumed an error of a factor of
1.5 in $\sin H_{\theta}$, our upper limit for the disk half-opening
angle for $\phi$~Per is half their adopted value.  In the case of
$\phi$~Per, the mass loss rate derived by \citet{Waters87} will be
overestimated by at least a factor of two, or more if the source is
not viewed edge-on.  Our upper limit on the half-opening angle of
17$^{\circ}$ for $\gamma$~Cas is consistent with the value adopted by
\citet{Waters87}.  However, based on the line profile classification
scheme established by \citet{Hanuschik96} we can conclude that
$\gamma$~Cas is not viewed edge-on and therefore its half-opening
angle is most likely less than 17$^{\circ}$.  In the case of the Be
star $\zeta$~Tau, for which the apparent ellipticity of the
circumstellar region has also been
detected~\citep{Quirrenbach97,Tycner04}, one can conclude that
$H_{\theta} \lesssim 9^{\circ}$.  It is possible that other Be stars
also have small disk opening angles.  To further test the generality
of this hypothesis, future interferometric observations should
concentrate on systems that are thought to be viewed at large
inclination angles~\citep[for example using H$\alpha$ profile
classification of][]{Hanuschik96}.

We can also compare our interferometrically determined opening angles
with those predicted by \citet{Stee03} who has calculated the disk
opening angles for a number of Be stars based on the flux in the
Brackett continuum near 2.2~$\mu$m using the SIMECA
code~\citep{Stee94,Stee95}.  The main characteristic of the SIMECA
model is that for a fixed disk density at the stellar photosphere, the
opening angle is directly proportional to the continuum flux, so that
a large flux in the $K$-band corresponds to a large opening angle.
For example, \citet{Stee03} obtained a value of 14$^{\circ}$ for the
full-opening angle of the disk of $\phi$~Per, which agrees with our
upper value of 16$^{\circ}$.  However, our upper limit for the
full-opening angle for $\gamma$~Cas is 34$^{\circ}$, which is
significantly less than the 54$^{\circ}$ obtained by \citet{Stee03}
for this star.  This discrepancy may be due to the fact that
\citet{Stee03} might have used apparent and not absolute magnitudes to
derive the disk-opening angles, in which case the values reported in
that study should be reevaluated.

\subsection{Disk Truncation}
\label{sec:disk_truncation}

\citet{Tycner04} showed that the H$\alpha$-emitting region of the
binary Be star $\zeta$~Tau was well within the estimated Roche lobe of
the primary, which suggests that the disk is truncated.  Another
example of a possible disk truncation was presented
by~\citet{Chesneau05} for $\alpha$~Ara, where the VLTI operating in
the $N$ band did not resolve any structure and therefore allowed the
authors to put an upper limit on the extent of the circumstellar
region.  Because this limit was smaller than the expected spatial
extent based on the models that were fitted to the Balmer emission
lines, the authors suggest that disk truncation by an unseen companion
might be occurring.

Evidence is accumulating which suggests that $\gamma$~Cas is a member
of a binary system.  \citet{Harmanec00} published the first orbital
radial velocity curve for $\gamma$~Cas based on observations collected
in the 628--672~nm range from 1993 to 2000.  Their analysis suggests
that this star is the primary component of a spectroscopic binary with
a period of $\approx 204$ days and an eccentricity of 0.26. This
result was confirmed by \citet{Mirosh02} using high-resolution
spectroscopic observations of the H$\alpha$ line obtained over a
similar time period. They report a periodic change of 205 days in this
line, which they also attribute to the binary system.  Although
\citet{Harmanec00} concludes that the secondary could be either a hot
compact object or a low luminosity late-type star, they estimate that
the separation between the components at periastron could be between
250 and 300~$R_{\sun}$.  The distance to $\gamma$~Cas is $188 \pm
20$~pc based on the {\it Hipparcos} satellite
measurements~\citep{Perryman97}, which places the separation between
the components at 6.2--7.4~mas.  Our best-fit value for the FWHM of
the major axis is $3.59\pm 0.04$ and therefore the circumstellar disk
is well within the orbit established by \citet{Harmanec00}.  However,
because the binary parameters of $\gamma$~Cas are not well established
this comparison is tentative.

The binary nature of $\phi$~Per is much better established since it
was recognized as a binary in the early
1900's~\citep{Cannon10,Ludendorff11}.  \citet{Poeckert81} documents
the development in the understanding of $\phi$~Per's binary nature
over the succeeding decades, and suggests that a subdwarf O star is
the secondary.  \citet{Gies98} detected the secondary in ultraviolet
spectra obtained with the Hubble Space Telescope and confirmed
Poeckert's prediction.  $\phi$~Per is now part of a group of four
candidate Be$+$sdO binaries, the others being HR
2142~\citep{Waters91}, 59~Cyg~\citep{Maintz03}, and
FY~CMa~\citep{Rivinius04}.  Be$+$sdO binaries may result from a
spin-up of the B star as a result of mass transfer from the progenitor
of the evolved subdwarf companion.

Estimates of the orbital parameters of $\phi$~Per are available in
\citet{Bozic95} and \citet{Gies98}. The semi-major axes of these two
orbital solutions range from approximately 230 to 310 R$_\odot$.
Using the {\it Hipparcos} distance of 220$^{+43}_{-31}$~pc for
$\phi$~Per~\citep{Perryman97} we expect an angular separation of
4.9--6.6 mas between the components of $\phi$~Per.  As we would
expect, this separation is larger than the angular radius of any of
the models of $\phi$~Per listed in Table~\ref{tab:best_fit}.  The
radial extent of the Gaussian model can be approximated with the FWHM
measure of 3.12$\pm$0.08~mas~(148$\pm$30~$R_{\sun}$) we obtain for
$\phi$~Per~(i.e., we are assuming that the radial extent is twice the
half-width of the Gaussian).  This disk radius is very close to the
178--204~$R_{\sun}$ Roche radius of the primary obtained by
\citet{Bozic95}.  This is most likely another example of a
H$\alpha$-emitting region that is close in size to the actual extent
of the circumstellar region, and this suggests that disk truncation is
occurring in this system.

We should note that in the case of $\phi$~Per the presence of a
truncated disk has been predicted.  \citet{Waters86} used the IRAS
observations of $\phi$~Per at 12, 25, and 60~$\mu$m to measure the IR
excess and constrained the density distribution of $\rho(R) = \rho_o
(r/R_{\star})^{-n}$ with $n=3.1$ for the disk model.  Because he
obtained a value of $n=2.4$ for two other Be stars, $\delta$~Cen and
$\chi$~Oph, which also happens to correspond to a velocity law that
has more gradual increase with distance~(something that is predicted
based on the H$\alpha$ spectroscopy), he argued that a density
distribution with $n=2.4$ might be more appropriate for $\phi$~Per,
which could be only satisfied if the disk model was truncated at $\sim
6.5 R_{\star}$~(or $\sim$46~$R_{\sun}$).  The apparent discrepancy
between our values for the size of the truncated disk might be related
to the different wavelength regimes of our studies, as well as to the
assumption made by \citet{Waters86} that the system is viewed pole-on,
whereas $i$ might be very close to 90$^{\circ}$.

\citet{Poeckert81} also predicted a second H$\alpha$-emitting disk
associated with the secondary component.  Although our $V^2$ data in
the H$\alpha$ channel does not have a very high SNR, we do not detect
any deviations from the elliptical Gaussian model that would suggest a
binary signature in the H$\alpha$ signal.  Therefore, we are forced to
conclude that if the secondary star does possess a H$\alpha$ emitting
disk it does not contribute significantly to the net H$\alpha$
emission line.  This is also consistent with the secondary's disk
radius of 6--8~$R_{\sun}$ estimated by \citet{Bozic95} based on the
peak separation of the He~{\footnotesize II} emission line.  Assuming
that the H$\alpha$ emission is proportional to the area of the disk,
as demonstrated by \citet{Tycner05}, we conclude that the H$\alpha$
emission from the disk of the secondary will contribute less than
$\sim$1\% to the total H$\alpha$ flux.

The direct detection of the secondary with the NPOI is unlikely as
\citet{Gies98} report a brightness ratio of 0.15~(or $\Delta$m of 2)
at 164.7~nm. The NPOI operates over a bandwidth of 550--850~nm and at
these wavelengths the magnitude difference between the two components
is likely to be much larger than 2.  A search for a binary
signature~(a sinusoidal variation) across the continuum channels in
both $\gamma$~Cas and $\phi$~Per yielded null results.  On the other
hand, a binary signature with a small $V^2$ amplitude~(less than
$\sim$3~\%) could not be ruled out.  In fact, it might be possible to
detect the signature of binarity in both stars at longer wavelengths
since $\gamma$~Cas may be associated with a cool companion, and
$\phi$~Per might contain large contributions from the circumstellar
disks of both components due to the free-free and free-bound emission.

\section{Conclusions}

We have successfully demonstrated the use of a narrowband filter in
the NPOI to increase the contrast between the H$\alpha$-emitting
material and the central star.  Our observations of two Be stars,
$\gamma$ Cas and $\phi$ Per, and our subsequent analysis have yielded
the following results:
\begin{enumerate}
\item We have demonstrated that the uniform disk or ring-like models
      are inconsistent with the observations of the circumstellar
      region of $\gamma$~Cas, whereas a Gaussian model is fully
      consistent with the data.  However, since the thermal structure
      of the circumstellar disks of Be stars can be quite complex, as
      suggested by recent models~\citep{Millar98,Carciofi06}, it might
      turn out that a more complicated brightness distributions may be
      required to fully describe these regions.
\item The circumstellar disk of $\gamma$~Cas appears to be consistent
      with the orbital parameters published in the literature.
      However, higher precision binary solutions are required to test
      for the possible disk truncation by the secondary.
\item Based on interferometric and spectroscopic data we have shown
      that the brightness distribution of the H$\alpha$-emitting
      circumstellar region of $\phi$~Per is best represented by a
      Gaussian distribution.
\item Our analysis supports the earlier prediction by \citet{Waters86}
      that the disk of $\phi$~Per is truncated due to the presence of
      an orbiting companion. However, the disk size we report is
      different than the value determined by \citet{Waters86}. This
      discrepancy is likely due to different wavelength regimes used
      in each study and/or due to the low inclination angle assumed by
      \citet{Waters86} for this star, whereas we obtain $i \gtrsim
      74^{\circ}$.
\end{enumerate}

A natural extension of the analysis presented in this study is the
direct comparison of interferometric data with theoretically predicted
interferometric observables, such as synthetic squared visibilities.
Future observations and modeling are planned that combine narrowband
interferometry with spectroscopy for several other Be stars,
especially those that possess weak H$\alpha$ emission.  The goal of
this work is to place greater constraints on the spatial extent and
physical properties of the circumstellar material.  These constraints
will allow various dynamical models to be tested with greater
certainty.

\bigskip
\small

The Navy Prototype Optical Interferometer is a joint project of the
Naval Research Laboratory and the US Naval Observatory, in cooperation
with Lowell Observatory, and is funded by the Office of Naval Research
and the Oceanographer of the Navy.  We would like to thank Susan
Strosahl and Dale Theiling for their contribution to the NPOI project
through nightly observing, and Brit O'Neill for the assistance with
the H$\alpha$ filter setup.  We would also like to acknowledge the
support of Jeff Munn for providing the observation planning software,
and Christian Hummel for the OYSTER reduction software.  We also thank
the anonymous referee for the useful comments on how to improve this
manuscript.  C.T. thanks Lowell Observatory for the generous time
allocation on the John~S.~Hall 1.1~m telescope and thanks Wes Lockwood
and Jeffrey Hall for supporting the Be star project on the
Solar-Stellar Spectrograph.  C.T. acknowledges that this work was
performed under a contract with the Jet Propulsion Laboratory~(JPL)
funded by NASA through the Michelson Fellowship Program, while being
employed by NVI, Inc.~at the US Naval Observatory.  JPL is managed for
NASA by the California Institute of Technology.  This research has
made use of the SIMBAD literature database, operated at CDS,
Strasbourg, France.


\newpage


\newpage

\begin{table}[htp]
\caption[]{\sc \small Observing Log for $\gamma$~Cas and $\phi$~Per}
\label{tab:obslog}
\begin{tabular}{lcccccc} \hline\hline
           &\multicolumn{3}{c}{$\gamma$~Cas (Number of Scans)}  & \multicolumn{3}{c}{$\phi$~Per (Number of Scans)} \\
\hspace{1.0cm} UT Date\hspace{1.2cm} &  AE--AC  &  AE--W7  &  AE--AW  & AE--AC  &  AE--W7  &  AE--AW  \\ \hline
   2004 Nov 3  \dotfill           & \ldots   &    2     &    2     & \ldots  &    2     &    2     \\
   2004 Nov 4  \dotfill           & \ldots   &    7     &    7     & \ldots  &    9     &    9     \\
   2004 Nov 5  \dotfill           & \ldots   &    7     &    7     & \ldots  &    8     &    8     \\
   2004 Nov 30 \dotfill           & \ldots   &  \ldots  &    5     & \ldots  &  \ldots  &    6     \\
   2004 Dec 1  \dotfill           & \ldots   &    2     &    2     & \ldots  &    8     &    8     \\
   2004 Dec 2$^{\ast}$  \dotfill  & \ldots   &    9     &    9     & \ldots  &   14     &   14     \\
   2004 Dec 3  \dotfill           & \ldots   &   10     &   10     & \ldots  &    7     &    7     \\
   2004 Dec 4  \dotfill           & \ldots   &    3     &    6     & \ldots  &    3     &    6     \\
   2004 Dec 10 \dotfill           &    9     &  \ldots  &    9     &    9    & \ldots   &    9     \\
   2004 Dec 11 \dotfill           &   11     &  \ldots  &   11     &   12    & \ldots   &   12     \\
   2004 Dec 12 \dotfill           &    8     &  \ldots  &    9     &   11    & \ldots   &   11     \\
   2004 Dec 19 \dotfill           &    9     &  \ldots  &    9     &    9    & \ldots   &   10     \\
   2004 Dec 23 \dotfill           &   12     &  \ldots  &   12     &   10    & \ldots   &   10     \\
\hline
\hspace{0.5cm} Total:             &   49     &   40     &   98     &   51    &   51     &   112    \\
\hline
\end{tabular}\\[0.9ex]
\parbox{6.0in}{\footnotesize \quad {\sc Note.} --- The AC--W7 baseline
was not accessible to $\gamma$~Cas~(see \S~\ref{sec:interferometry})
and therefore the AC station was not used until after 2004~Dec~4, when
W7 station came off line due to gate valve work.  $^{\ast}$ -- The
narrowband H$\alpha$ filter was not used on 2004 Dec 2.}
\end{table}

\newpage

\begin{table}[htp]
\caption[]{\sc \small Best-Fit Model Parameters for $\gamma$~Cas and $\phi$~Per}
\label{tab:best_fit}
\small
\begin{tabular}{llcccccc} \hline\hline
                      &             & $\theta         $ &                 &                  &    $\phi$      &                 &            \\
 Be Star \hspace{5mm} & Model       &     (mas)         &    $\epsilon$   &       $r$        &    (deg)       &    $c_{\rm p}$  & $\chi^2_{\nu}$ \\
\hline
$\gamma$~Cas \dotfill & UD          &  $4.72 \pm 0.03$  &     $\ldots$    & $0.89 \pm 0.02$  & $29.6\pm 2.5$  & $0.53 \pm 0.01$ & 6.0  \\
                      & UD$^{\ast}$ &  $6.49 \pm 0.06$  &     $\ldots$    & $0.52 \pm 0.03$  & $34.7\pm 1.5$  & $0.59 \pm 0.01$ & 1.6  \\
                      & UR          &  $7.91 \pm 0.09$  & $0.04$ -- $0.90$& $0.58 \pm 0.03$  &   $5$ -- $40$  & $0.58 \pm 0.13$ & 5.7  \\
                      & UR$^{\ast}$ &  $5.60 \pm 0.53$  & $0.77 \pm 0.16$ & $0.52 \pm 0.03$  & $36.0\pm 1.5$  & $0.67 \pm 0.01$ & 1.7  \\
                      & GD          &  $3.59 \pm 0.04$  &     $\ldots$    & $0.58 \pm 0.03$  & $31.2\pm 1.2$  & $0.51 \pm 0.01$ & 1.4  \\
                      & GD$^{\ast}$ &  $3.64 \pm 0.05$  &     $\ldots$    & $0.56 \pm 0.03$  & $30.6\pm 1.4$  & $0.51 \pm 0.01$ & 1.4  \\
$\phi$~Per   \dotfill & UD          &  $5.43 \pm 0.13$  &     $\ldots$    & $0.27 \pm 0.01$  & $-61.5\pm 0.6$ & $0.52 \pm 0.01$ & 1.4  \\
                      & UR          &  $5.23 \pm 0.33$  & $0.58 \pm 0.12$ & $0.26 \pm 0.01$  & $-61.7\pm 0.6$ & $0.62 \pm 0.01$ & 1.4  \\
                      & GD          &  $2.89 \pm 0.09$  &     $\ldots$    & $0.27 \pm 0.01$  & $-61.5\pm 0.6$ & $0.39 \pm 0.02$ & 1.4  \\
\hline
\end{tabular}\\[0.9ex]
\parbox{6.0in}{\footnotesize \quad {\sc Note.} --- For the UD and UR
models $\theta$ corresponds to the angular diameter of the major axis,
and for GD model it corresponds to the FWHM of the major axis of the
elliptical Gaussian.  $^{\ast}$ - models fitted to data from the
shortest two baselines only.}
\end{table}

\begin{table}[htp]
\caption[]{\sc \small Circumstellar Regions of $\gamma$~Cas and
$\phi$~Per}
\label{tab:parameters}
\small
\begin{tabular}{lccc} \hline\hline
        Description      &        Symbol         &          $\gamma$~Cas      &   $\phi$~Per \\ 
\hline
Disk size (mas)          &  $\theta_{\rm GD}$    &       3.59 $\pm$ 0.04      &  2.89 $\pm$ 0.09  \\
Axial Ratio              &          $r$          &       0.58 $\pm$ 0.03      &  0.27 $\pm$ 0.01  \\
P.A. of major axis       &       $\phi$          & $31\fdg2 \pm 1\fdg2$       &  $-61\fdg5 \pm 0\fdg6$    \\
Inclination              &       $i$             &    $\gtrsim 55^{\circ}$    & $\gtrsim 74^{\circ}$  \\
Half-opening angle       &     $H_{\theta}$      &    $\lesssim  17^{\circ}$  & $\lesssim  8^{\circ}$  \\
\hline
\end{tabular}\\
\end{table}


\end{document}